\documentclass[preprint2]{aastex}

\shorttitle{Andromeda~XIX}
\shortauthors{Cusano et al.}

\begin{document}

\title{DWARF SPHEROIDAL SATELLITES OF M31: I. VARIABLE STARS AND STELLAR POPULATIONS IN ANDROMEDA~XIX\altaffilmark{*}}

\author{FELICE CUSANO\altaffilmark{1}, GISELLA CLEMENTINI\altaffilmark{1}, ALESSIA GAROFALO\altaffilmark{1,2}, 
MICHELE CIGNONI\altaffilmark{2}, LUCIANA FEDERICI\altaffilmark{1}, MARCELLA MARCONI\altaffilmark{3},  
ILARIA MUSELLA\altaffilmark{3}, VINCENZO RIPEPI\altaffilmark{3},
KONSTANTINA BOUTSIA\altaffilmark{4}, MARCO FUMANA\altaffilmark{5}, STEFANO GALLOZZI\altaffilmark{4}, VINCENZO TESTA\altaffilmark{4}}

\affil{$^1$INAF- Osservatorio Astronomico di Bologna, Via Ranzani 1, I - 40127 Bologna, Italy}
\email{felice.cusano@oabo.inaf.it, gisella.clementini@oabo.inaf.it, luciana.federici@oabo.inaf.it}

\affil{$^2$Dipartimento di Astronomia, Universit\`a di Bologna,  Via Ranzani 1, I - 40127 Bologna, Italy}
\email{alessia.garofalo@studio.unibo.it, michele.cignoni@unibo.it}

\affil{$^3$INAF- Osservatorio Astronomico di Capodimonte, Salita Moiariello 16, 
I - 80131 Napoli, Italy}
\email{marcella@na.astro.it, ilaria@na.astro.it, ripepi@na.astro.it}

\affil{$^4$INAF- Osservatorio Astronomico di Roma, Via di Frascati 33
00040 Monte Porzio Catone, Italy}
\email{konstantina.boutsia@oa-roma.inaf.it, stefano.gallozzi@oa-roma.inaf.it, vincenzo.testa@oa-roma.inaf.it}

\affil{$^5$INAF - IASF Milano, Via E. Bassini 15, I-20133 Milano, Italy}
\email{fumana@lambrate.inaf.it}

\altaffiltext{*}{Based on data collected  with Large Binocular Cameras at the Large Binocular Telescope}

\begin{abstract}
We present $B,V$ time-series  photometry of \object{Andromeda~XIX} (And~XIX), the most extended (half-light radius  of  $6.2^{\prime}$) of   Andromeda's  
%investigated the  variable stellar population of \object{Andromeda~XIX}, a 
dwarf spheroidal companions,  that we observed with the Large Binocular Cameras at the Large Binocular Telescope.
We surveyed a $23^{\prime} \times 23^{\prime}$ area centered on And~XIX and present the deepest color magnitude diagram  (CMD) 
ever obtained for this
galaxy,  reaching,  at $V \sim$ 26.3 mag, about one magnitude below the horizontal branch (HB).  The CMD shows a prominent and slightly 
widened red giant branch, along with a predominantly red HB, which, however, extends to the blue to significantly  populate the classical 
instability strip.
 We have identified  39 pulsating variable stars, of which 31 are of RR Lyrae type and 8 are Anomalous Cepheids (ACs). 
Twelve of the RR Lyrae variables and 3 of the ACs are located  within And~XIX's half light radius.
%, thus indicating that the
% galaxy experienced  a t $>$ 10 Gyr old star formation event, possibly along with an intermediate age stellar generation.  
The average period of the fundamental  mode RR Lyrae stars ($\langle Pab \rangle$ = 0.62 d, $\sigma$= 0.03 d) and the period-amplitude diagram 
qualify And~XIX as an  Oosterhoff-Intermediate system.
From the average luminosity of the RR Lyrae stars ($\langle V (RR)  \rangle$ = 25.34 mag, $\sigma$= 0.10 mag) %, average on 26 stars)
 we determine a distance modulus of (m-M)$_0$=$24.52\pm0.23$ mag in a scale where the distance to the Large Magellanic Cloud (LMC) is $18.5\pm0.1$ mag.
 The ACs follow a well defined  Period-Wesenheit ($PW$) relation that appears to be in very good agreement with the  $PW$ relationship defined by the ACs in the LMC. %   recently published in the literature.
\end{abstract}


\section{INTRODUCTION}\label{sec:intro}
The number of satellites known to surround the Andromeda spiral galaxy  (M31)  has increased  dramatically in the last 
few years thanks to the systematic imaging  of the
M31 halo being  carried out 
%mainly  with the Canada-France-Hawaii Telescope 
 by the Pan-Andromeda Archaeological Survey \citep[PAndAS; ][]{mart06,mart09,iba07,irw08, mac08, mac09,rich11},   
the Sloan Digital Sky Survey  \citep[SDSS; ][]{zuck04, zuck07,slat2011, bell2011}, and, lately, the Panoramic Survey Telescope and Rapid 
Response System 1 survey \citep[Pan-STARRS1; ][]{mart13}.
The latest census of the M31 companions currently counts  31 dwarf spheroidal galaxies (dSphs) whose  luminosities range 
from 10$^4$ to 10$^8 L_{\odot}$. This number is expected to further grow as new diffuse stellar systems are being discovered in 
the M31 halo \citep[e.g.  PAndAS-48;][]{mack13}   whose actual nature, whether extended globular clusters (GCs) or ultra-faint dwarfs (UFDs) 
like those discovered around the Milky Way (MW), remains to be established. 
In the framework of the hierarchic  formation of structures the dSph satellites we observe today around M31 may be the  survivors 
of Andromeda's building process. Their stellar content can thus provide insight to reconstruct the star formation history (SFH) 
and the merging episodes that led to the early assembling of the M31 halo. 
%The most direct way  understanding the formation history of a galaxy over the whole Hubble time is the analysis of colorÐmagnitude diagrams (CMDs) deep enough to reach the main-%sequence turnoff (TO) of the oldest populations.
%The SFHs through accurate synthetic modeling of their CMDs
The synthetic modeling of deep color magnitude diagrams (CMDs) represents the most direct way for understanding the formation history 
of a galaxy. 
However,  the CMDs currently available for the M31 dSphs  generally sample only the brightest stars 
\citep[e.g. ][]{zuck04,mac08,rich11,bell2011},  and even  when Hubble Space Telescope (HST)  data are available  
\citep{pri02,pri04,pri05, manc08, yang12} they  only  reach slightly below the HB, as observing  
the main sequence turn-off (MSTO) of the oldest stellar populations  at the distance of M31 
[$(m- M)_0= 24.47 \pm 0.07$ mag, D= 783 kpc, \citealt{mac12}, or  $(m- M)_0= 24.42 \pm 0.06$ mag, D= 766 kpc, \citealt{fede12}]requires of the order of tens of HST orbits. 

The pulsating variable stars are a powerful  alternative tool to  investigate the different stellar generations occurred 
in the M31  dSphs, as Classical Cepheids can be used to trace  the young stars (typical ages ranging from a few to a few hundred Myr),  
%the Anomalous Cepheids (ACs)  trace the intermediate age  (t$\sim$ 1-2 Gyr) stellar component, and, most importantly,  
and the RR Lyrae  stars, which are comparably old but about 3 magnitude brighter than the MSTO, 
can allow to unravel the oldest stars born more than 10 Gyr ago.  
Using characteristics   like the mean period of the fundamental-mode
  (RRab) and first-overtone (RRc)  pulsators   and the  ratio of  number of RRc ($N_c$) to  total number of RR Lyrae stars ($N_{ab+c}$) 
  ($f_c = N_c/N_{ab+c}$) the MW field and cluster RR Lyrae are divided into two different  
groups \citep{oos39}:  Oosterhoff type I (Oo~I) clusters  have $\langle P_{\rm ab} \rangle$ $\simeq$ 0.55 d, 
$\langle P_{\rm c }\rangle$ $\simeq$ 0.32 d,  and  $f_c \sim 0.17$;  and 
 Oosterhoff type II (Oo~II)  clusters have $\langle P_{\rm ab} \rangle$ $\simeq$ 0.65 d, $\langle P_{\rm c}\rangle$ $\simeq$ 0.37 d, 
and   $f_c \sim 0.44$. % ( e.g. %$\citep{cle01}
 %Contreras Ramos et al. 2013).
%Differences in between the two groups are also found in the ratio between the number of RRc to RRab stars that 
%is 0.22 in the Oo-I clusters and 0.48 in the Oo-II clusters \citep{cle01}.
Different Oosterhoff types also imply slightly different ages and metallicities \citep{van93}, as  Oo~II   clusters are  more metal-poor 
and older than Oo~I systems. 
Recent studies \citep[see e.g.][ for a review]{cat09} confirmed the Oosterhoff dichotomy to occur not only among the MW GCs, 
but also for field variable stars in the MW halo.
%   and also the  Oo-II are slightly older than Oo-I.
%     \citet{cat09} in a review of all the  GCs of the MW
 %     confirmed the Oosterhoff dichotomy  for the halo. 
On the other hand, the  separation in Oosterhoff groups seems to be  a characteristic 
of the MW, since the  dSphs around our galaxy,  as well as their respective GCs, have 
0.58 $\leq \langle {\rm P}_{\rm ab}\rangle \leq$ 0.62 d, and Oo-Intermediate (Oo-Int)
properties \citep[][ and references therein]{cat09,cle10}. %, 0.58 $\leq \langle {\rm P}_{\rm ab}\rangle \leq$ 0.62 days).
The characterization  of the RR Lyrae population in M31 and in its companions is still at an early stage.
 In a deep survey of
the M31 halo, \citet{bro04} identified 55 RR Lyrae stars
(29  RRab, 25 RRc, and
one double-mode, RRd, pulsator) in a $3.5'\times3.7'$ field along
the southeast minor axis of the galaxy. Based on their pulsation
properties, \citet{bro04} concluded that unlike in the MW the old population in the M31 halo 
has Oo-Int properties. 
%and
%does not conform to the subdivision in the Oosterhoff  types  followed by the
%RR Lyrae stars in the MW field and globular clusters.
However, a different conclusion was reached by \citet{sar09}
 who identified 681 RR Lyrae variables (555 RRab's, and
126 RRc's) based on the HST/Advanced Camera for Surveys (ACS) observations of two fields
near M32, at a projected distance between 4 and 6 kpc from the
center of M31, and concluded that these M31 fields have Oo-I
properties. A total of 108 RR Lyrae stars were identified by \citet{jef11} in six
HST/ACS ultra-deep fields located in the disk, stream and halo  of M31, showing that the RR Lyrae population appears  mostly to be 
of the Oo-I  and Oo-Int types.
Of the M31 globular clusters only  two had the RR Lyrae stars fully characterized,   
B514  was found to have Oo-Int properties  \citep{cle09} and 
G11 to be an Oo~II GC \citep{contr13}.
Six of the  M31 dSphs have been analysed so far for variability %,  based on Wide Field Planetary Camera 2 onboard the Hubble Space Telescope data 
 and RR Lyrae stars 
have been identified in all of them \citep{pri02, pri04, pri05, manc08, yang12}.  
According to these studies 
%
%Pritzl et al. (2004, 2005) based on  Wide Field Planetary Camera 2 data of  HST programs XXX. More recently  
%In M31, the systematic study of its variable stars population is fairly recent. Only five ÒclassicalÓ dSph galaxies have been analysed so far (Pritzl XXX, Sarajedini private comunication?) and, 
 all three Oosterhoff types (Oo I, Oo II, and Oo-Int) seem to be present among the M31 satellites.
% On the other hand, according to our results based on the study of the Oosterhoff dichotomy in GCs (Contreras et al. 2008, Clementini et al. 2009, Contreras et al. in preparation), the %Oosterhoff type II seems to be lacking among the M31 globular clusters. Finally, the situation is completely unknown regarding UFD galaxies, since no studies have been already done on %these objects.
These previous studies of  variables in the M31 dSphs are based on  Wide Field Planetary Camera 2  onboard the HST data. 
However, the HB of the M31 satellites can easily be reached from the ground with  8-10 m class telescopes.  
%Besides being less oversubscribed, 
The ground-based facilities usually allow  to cover areas significantly larger than the half-light radius ($r_h$) of the M31 
satellites, which often are rather extended, thus attaining much complete and statistically significant samples.  
We have obtained multi-band photometry of a sample of the M31 dSph satellites \citep[see][]{cle11} using the Large Binocular Telescope (LBT) 
and the Gran Telescopio Canarias (GTC) and in this paper, we present results from our study of the stellar population and variable stars 
in  Andromeda~XIX \citep{mac08}, the most extended of Andromeda's dSph companions. %
%
%
%
%
%Pritzl et al. in three different works studied the variable star population
%of four M31 dSph satellites, nominal And~VI \citep{pri02}, And~II \citep{pri04}, And~I andAnd~III \citep{pri05}.
%They found that three of these dSphs have a mean period of the RR Lyrae stars consistent with an Oo-Int classification
% with 
%$\left\langle\rm{P_{ab}}\right\rangle=0.571$, 0.588 and 0.621 days 
%for And~II, And~VI and And~I, respectively. And~III based on the mean period $\left\langle\rm{P_{ab}}\right\rangle=0.648$ days 
%can be classified as an Oo-II object. Nevertheless, the Amplitude-Period diagram of the RR Lyrae stars in these four M31 companions
%follow the relation defined by the RR Lyrae stars found by \citet{bro04}. 
%
%
%Aggiungere Clementini et al. 2009 (B514) e Contreras Ramos et al. 2012 (G11).
%
%While these previous studies used HST archive data, the horizontal branch of the M31 satellites can be easily reached from the ground with a 4-8 meter class telescope  
% 
%In this context we started a survey to characterize the variable stellar population of 
%the M31 companions using several telescope \citep[see][]{cle11} among which the Large Binocular Telescope (LBT)
%using the Large Binocular Camera (LBC).
%The first object observed in the LBT  survey is the dSph galaxy 

Andromeda~XIX \citep[And~XIX, R.A.$=00^{h}19^{m}32.1^{s}$,
DEC. $=+35^{\circ}02^{\shortmid}37.1^{\shortparallel}$, J2000.0;
 l=$115.6^{\circ}$, b=$-27.4^{\circ}$;][]{mac08} 
was discovered by \citet{mac08} in a photometric survey of the southwestern quadrant of M31 performed with the Megaprime camera of 
the Canada-France-Hawaii Telescope (CFHT). The galaxy 
is located at a projected distance  of $\sim$ 120 kpc \citep{con12} from the center of M31. The discovery data show 
a steep red giant branch (RGB) but do not reach deep enough to sample the galaxy HB.   
\citet{mac08} estimate for And~XIX an heliocentric distance  of 933 kpc using the luminosity of the RGB tip, 
however a closer distance of 821 kpc is derived by \citet{con12} applying to  the same data a bayesian approach 
to estimate the luminosity of the RGB tip. 
With  a half-light radius of $r_h=6.2^{\prime}$ (corresponding to a linear extension of $\sim$ 1.7 or 1.5 kpc whether \citealt{mac08} or
\citealt{con12}  distance estimate is adopted) And~XIX is the largest of Andromeda's dSph companions as well as the 
most extended of the Local Group (LG) satellites.
In the stellar density map presented by \citet{mac09} an overdensity of stars named Southwest Cloud seems
to connect And~XIX with the halo of M31,  providing hints of a possible interaction between And~XIX and Andromeda itself. 
\citet{col13} investigated spectroscopically  27 stars  located around the RGB of And~XIX  measuring a small velocity dispersion when compared
 to the galaxy radial extent. The authors attributed this ``cold'' velocity dispersion to the tidal interaction with M31.
From the spectra they also derived an average value of the metallicity  [Fe/H]=$-1.8\pm0.3$ dex 
 using the equivalent width of the calcium triplet and the Starkenburg et al. (2010) method, 
which is consistent with the value of [Fe/H]=$-1.9\pm0.1$ dex
 found photometrically by \citet{mac08}  by isochrone-fitting of the galaxy CMD.

The paper is organized as follow: observations, data reduction and calibration of And~XIX photometry are presented in Section 2. 
Results on the identification and characterization of the variable stars, 
the catalog of light curves, and the Oosterhoff classification of the RR Lyrae stars are discussed in  Sections 3 and 4. 
The distance to And~XIX derived from  the  RR Lyrae stars is  presented in Section 5.
The galaxy CMD is presented in Section 6. Properties and classification of the variable stars above the HB are discussed in Section 7.
In Section 8 an estimate of the contamination from the halo of M31 is given. The discussion on the spatial distribution 
of the And~XIX's stars is presented in Section 9. 
In Section 10 we give an interpretation of the CMD using stellar isochrones and evolutionary tracks. 
Finally, a summary of the main results is presented in Section 11.
%Lv(L$_\odot$)=6.49*10^5 

\section{OBSERVATIONS AND DATA REDUCTIONS}

Time series $B$, $V$  photometry of  And~XIX  was obtained in fall 2010 using the Large Binocular Cameras 
(LBC\footnote{See $http://lbc.oa-roma.inaf.it/$}) mounted at the foci of the  LBT.  
Each LBC consists of an array of 4 CCDs with total field of view (FoV) of  $\sim 23'\times23'$ and pixel scale of   0.225$``$/pixel. 
The two LBCs are optimized for the blue and red portion of the
visible spectrum.
The  $B$ exposures were obtained with the Blue LBC, whereas the $V$ exposures were obtained with the Red LBC.
We obtained 44 $B$ and 31 $V$ images each corresponding to a 420$^s$ exposure, for  total exposure times of  18480$^s$ and 13020$^s$ 
  in $B$ and $V$, respectively.
The log of And~XIX observations is  provided in Table~\ref{t:0}.

\begin{table*}
\begin{center}
\caption[]{Log of And~XIX observations}
\label{t:0}
\begin{tabular}{l c c c c }
\hline
\hline
\noalign{\smallskip}
   Dates                 & {\rm Filter}  & N   & Exposure time &  {\rm Seeing (FWHM)}    \\
 	                 &		 &     &	      (s)       &    {\rm (arcsec)}\\
\noalign{\smallskip}
\hline
\noalign{\smallskip}
  October   8, 2010     &   $B$      & 2  & 420    &  1.5\\
  December  1-3, 2010  &   $B$      & 42 & 420    & 0.7  \\ 
                         &             &   &        &        \\  
  October   8-11, 2010   &   $V$      &  6 &   420  &  1.3-2.0 \\
  December  1, 2010   &   $V$      & 25 &  420  &  0.8   \\
\hline
	  %%%%%%%%%%%%%%  \end{array}
	 %%%%%%%%%%%   $$
\end{tabular}
\end{center}
\normalsize
\end{table*}

Each image was pre-reduced (bias-subtracted, flat-fielded and astrometrized)  by the LBT team
through the LBC dedicated pipeline. 
PSF photometry was then performed using
  the \texttt{DAOPHOT- ALLSTAR-ALLFRAME} package  \citep{ste87,ste94}. 
  The  alignment of the images was performed using  \texttt{DAOMATCH}, one 
  of the routines in the  \texttt{DAOPHOT} package, whereas  \texttt{DAOMASTER}
  \citep{ste92} was used to match point sources.
  The absolute photometric calibration to Johnson  $B$ and $V$ was performed using stars in the scientific fields, 
 for which calibrated photometry is available 
  from previous studies.
As a first step,  we  cross-matched our photometric catalogs composed by 
  sources in the four CCDs  with the SDSS catalog  
  \citep{aba09}. The SDSS catalog was queried only for objects flagged as stars by the reduction package, 
   and with good quality of the observations.
A total of 1121 stars were found to be common between the two catalogs.
 The calibration equations %of Lupton (2005) 
  available at \textit{http://www.sdss.org/dr4/algorithms/sdssUBVRI Transform.html}
were used to convert the $g$ and $r$ magnitudes of the SDSS stars  to  the Johnson  $B$ and $V$ magnitudes. 
The parameters of the photometric calibration were finally derived 
 fitting the data to the equations $B-b$= c$_B$+m$_B\times$($b-v$) and 
 $V-v$=c$_V$+m$_V\times$($b-v$), where $B$ and $V$ are the standard Johnson magnitudes of  the SDSS stars,  
and $b$ and $v$ are the  instrumental magnitudes in our  LBC catalog. 
The fit was performed using a 3$\sigma$ clipping rejection algorithm. A total of 512 stars were used in the final fit,
 with magnitudes ranging from 15.7 to 24.3 mag in $B$,  and  from 0.2 to  1.7 mag in the $B-V$ 
color\footnote{The coefficients of the calibration equations are: c$_B$=27.532, m$_B=-$0.137, c$_V$=27.530, m$_V= -$0.0633}.
The final r.m.s. of the fit is of 0.04 mag both in $B$ and $V$.

\section{IDENTIFICATION OF THE VARIABLE STARS}\label{sec:var}

The first step in the identification of variable stars 
 was to search for objects with higher values 
 of the variability index computed in \texttt{DAOMASTER} \citep{ste94}.
We then also checked for variability all the stars with colors and magnitudes falling within the edges of the 
RR Lyrae and AC instability strips (ISs), according to the definition by \citet{mar04}.
 The final list of candidate variable stars consisted of $\sim$ 300 sources that all were 
inspected visually 
%
 %A sample of 300 candidate variable stars was thus collected. 
 %The $B$ light curves of the stars in this sample were analyzed
 using the Graphical Analyzer of Time Series package (GRaTIS),  
 custom software developed at the Bologna Observatory by P. Montegriffo
(see, e.g., \citealt{clm00}).
GRaTiS uses both the Lomb periodogram \citep{lom76,sca82} 
and the best fit of the data with a truncated Fourier series 
\citep{bar63}. 
The majority of the candidate variables were identified in both photometric bands, however,  
in few cases we have a reliable light curve only in the $B$ band. 
The period search was performed first on the $B$ band for which we have a larger number of epochs (44 phase-points). %  and then 
%used to phase the $V$ data. % for which we have less epochs.
 Final periods were then derived through an iterative procedure between the two
photometric bands. The  depth and sampling 
of the $B$ data is such that we were able to detect variables as faint as $B \sim 26$ mag, 
with periods ranging from a few hours to several days. 
A total  of 39 bona-fide variable stars were identified, of which 31 are RR Lyrae stars 
(23 RRab's,  and 8 RRc's) lying on the galaxy HB, 
 and  8 were classified as  ACs on the basis of the luminosities about 1-1.5 mag  brighter than the HB level, and 
 the comparison with the AC IS, theoretical isochrones,  and the Period-Wesenheit ($PW$) relation  
 (see Sects.~\ref{sec:ACS}, and ~\ref{sec:recent}).
  Among them are 5 RR Lyrae stars and  1 AC for which we have 
reliable data only in the $B$ band.  
 Identification and properties of the confirmed variable stars  are 
  summarized in Table~\ref{t:1}. 
Column 1 gives the star identifier. We assigned to the variables an increasing number starting from the galaxy center, 
for which we adopted \cite{mac08}'s coordinates.
Columns 2 and 3 provide the right ascension and declination (J2000 epoch), respectively. 
These coordinates were obtained from our astrometrized catalogs. Column 4 gives the type of variability.
Columns 5 and 6 list the pulsation period and the Heliocentric Julian Day (HJD) of maximum light, respectively. 
Columns 7 and 8 give the intensity-averaged mean $B$ and $V$ magnitudes, 
while Columns 9 and 10 list the corresponding amplitudes of the light variation. 
 Light curves for the 39 stars are shown in Figure~1. To calibrate the photometry of  stars
that only have $B$ light-curves we simply added to the instrumental photometry the zero point c$_B$ 
of the $B$ calibration equation.
\begin{figure*}\label{fig:lca}
\centering
\includegraphics[width=14cm,height=16cm]{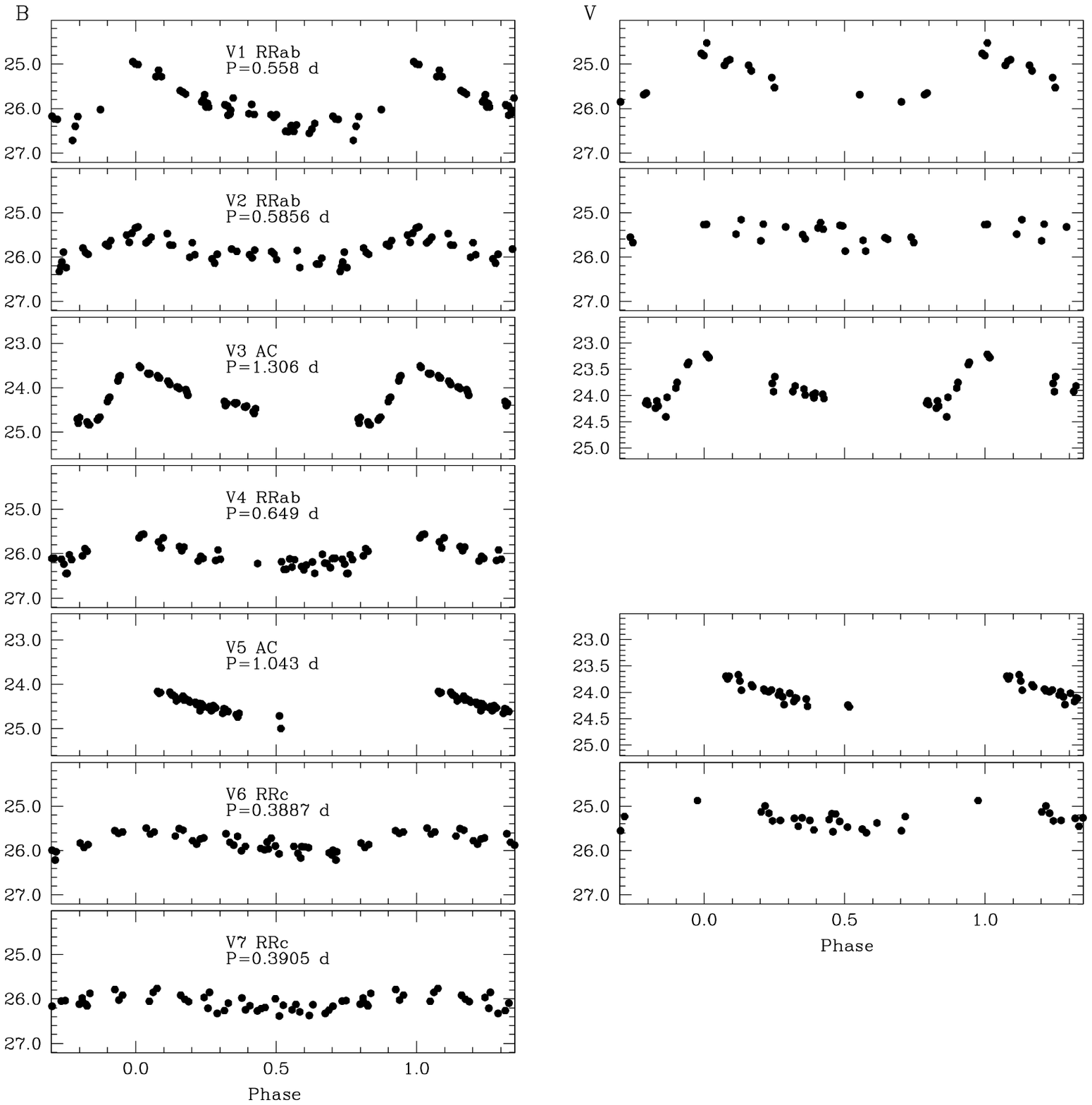}
\caption{$B$ (\textit{left panels}) and $V$ (\textit{right panels}) light curves of the variable stars identified  in And~XIX. 
Stars are ordered with increasing distance from the galaxy  center, for which we adopted the \cite{mac08} coordinates.
Typical internal errors for the single-epoch data are in the range of 0.04 to 0.12 mag in $B$, and of 0.08 to 0.19 mag in $V$.}
\end{figure*}
\begin{figure*}
\centering
\figurenum{1}
\includegraphics[width=14cm,height=16cm]{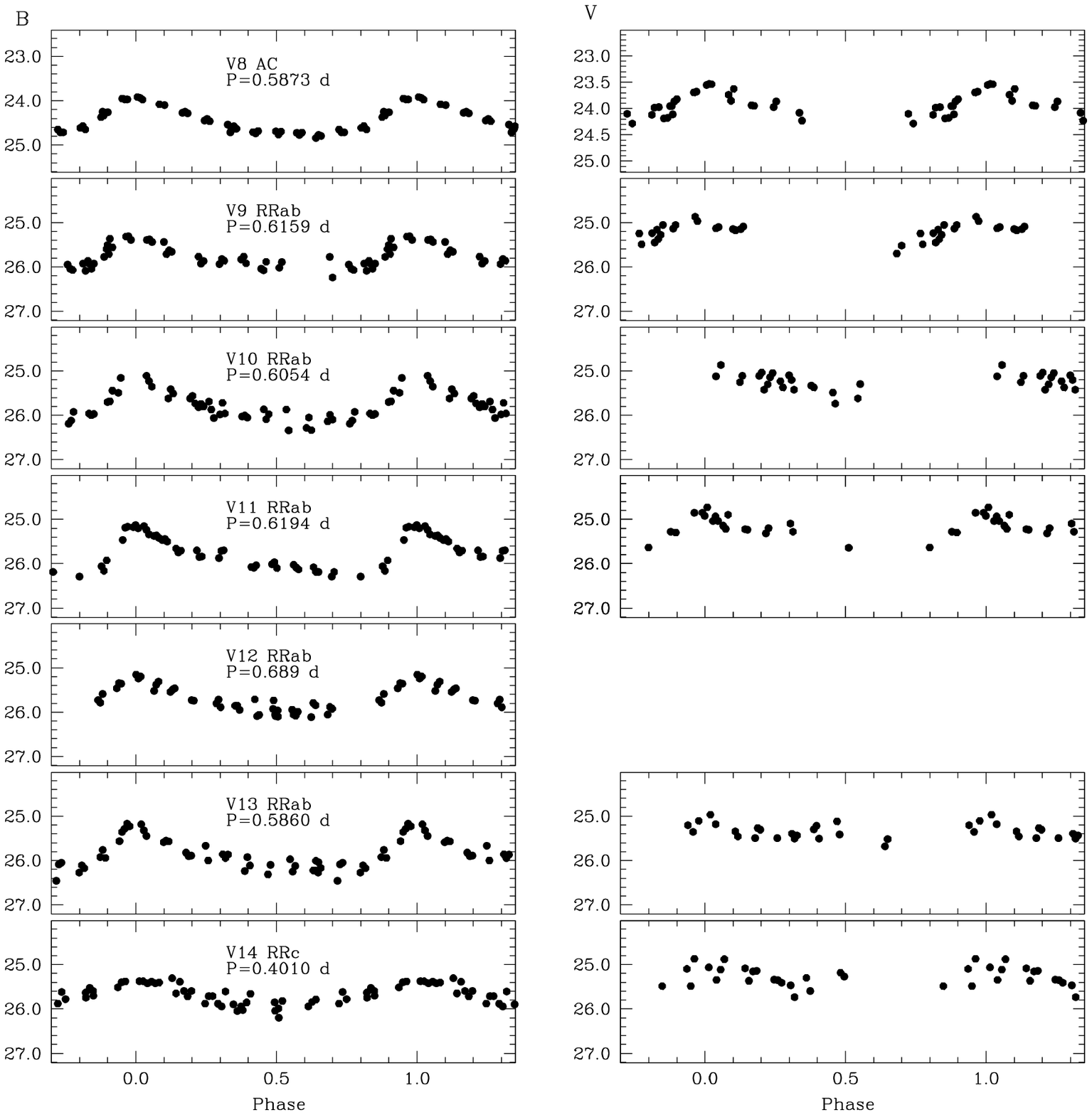}
\caption{ continued --}
\end{figure*}
\begin{figure*}
\centering
\figurenum{1}
\includegraphics[width=14cm,height=16cm]{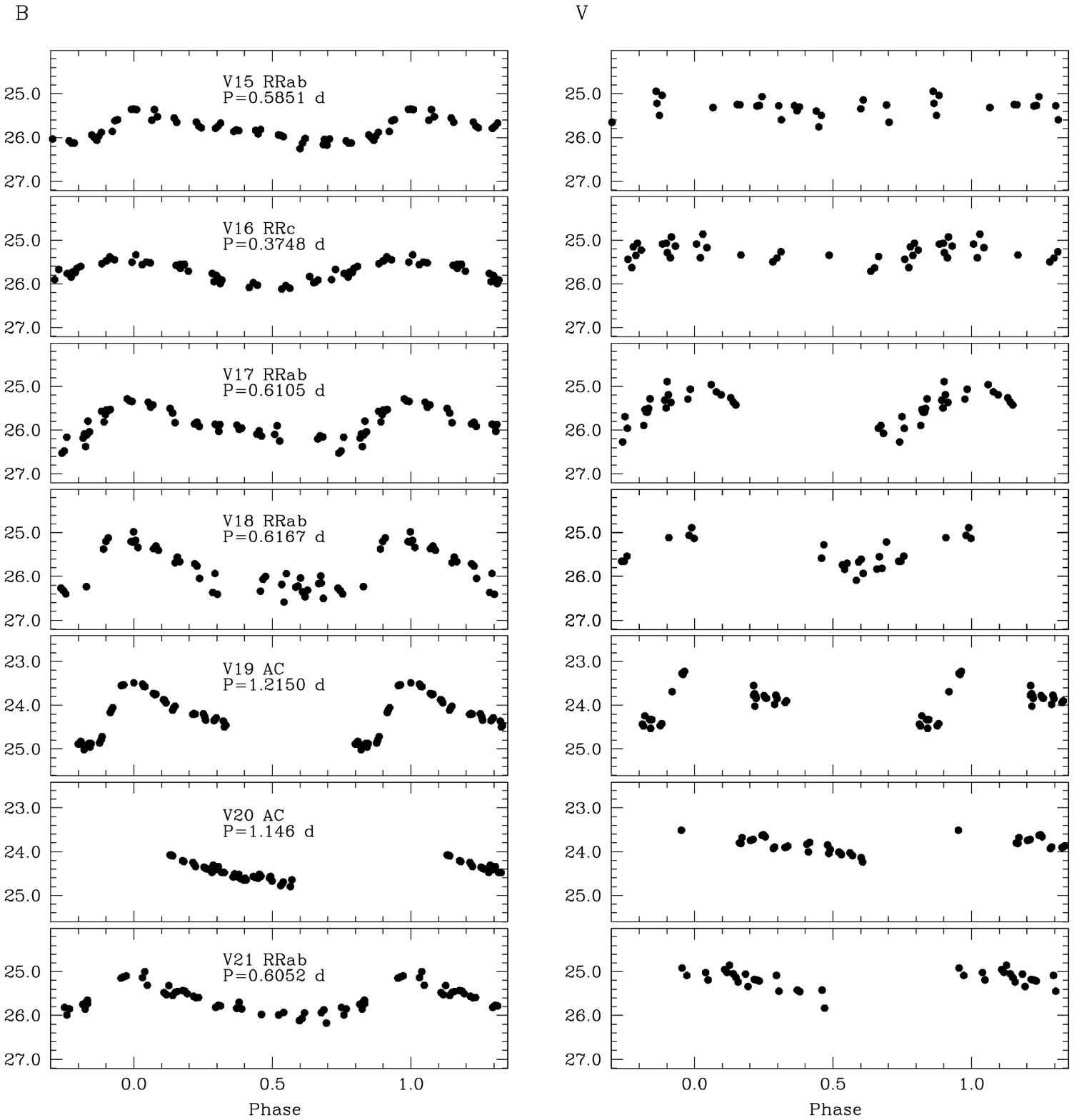}
\caption{ continued --}
\end{figure*}
\begin{figure*}
\centering
\figurenum{1}
\includegraphics[width=14cm,height=16cm]{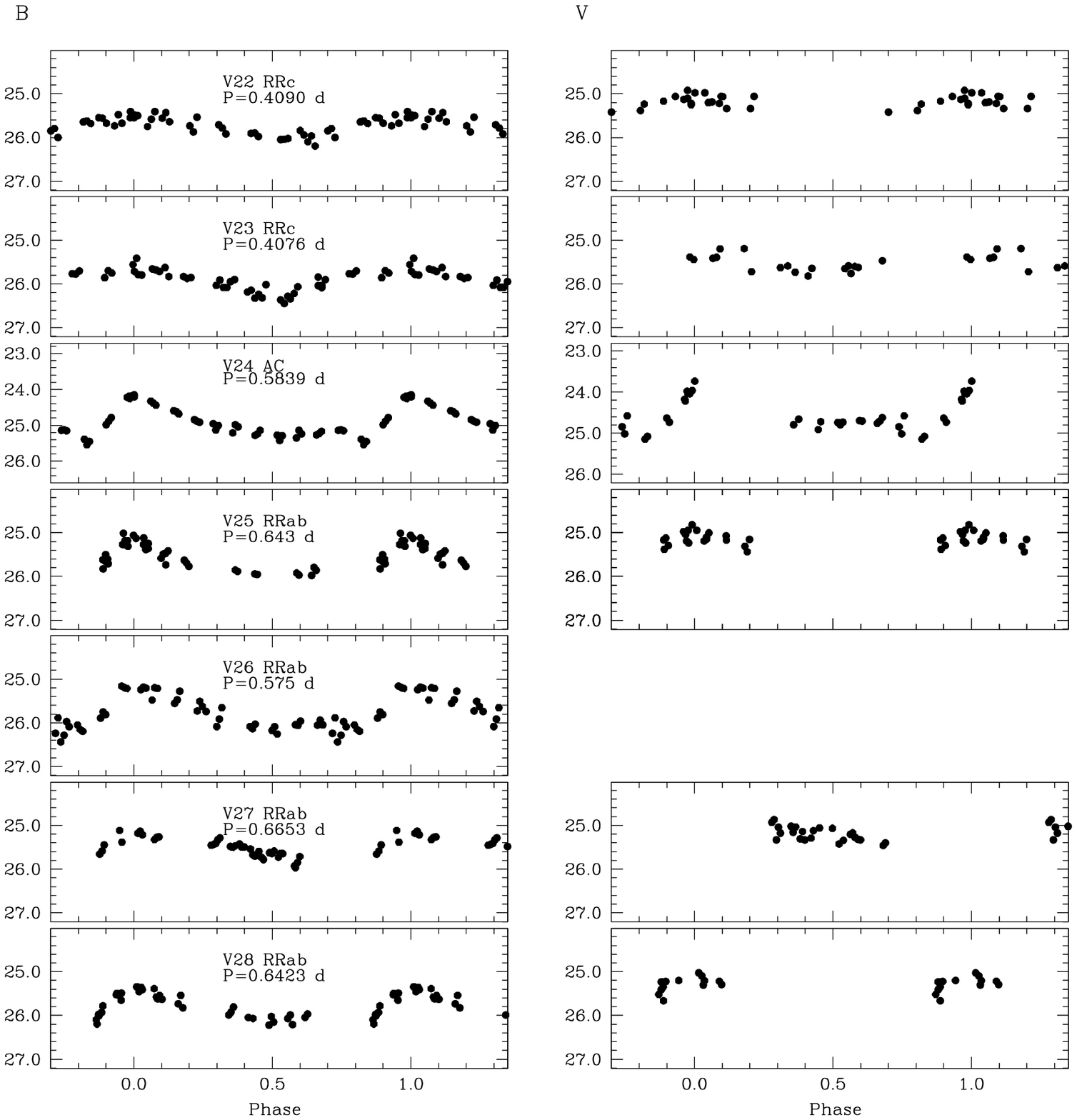}
\caption{ continued --}
\end{figure*}
\begin{figure*}
\centering
\figurenum{1}
\includegraphics[width=14cm,height=16cm]{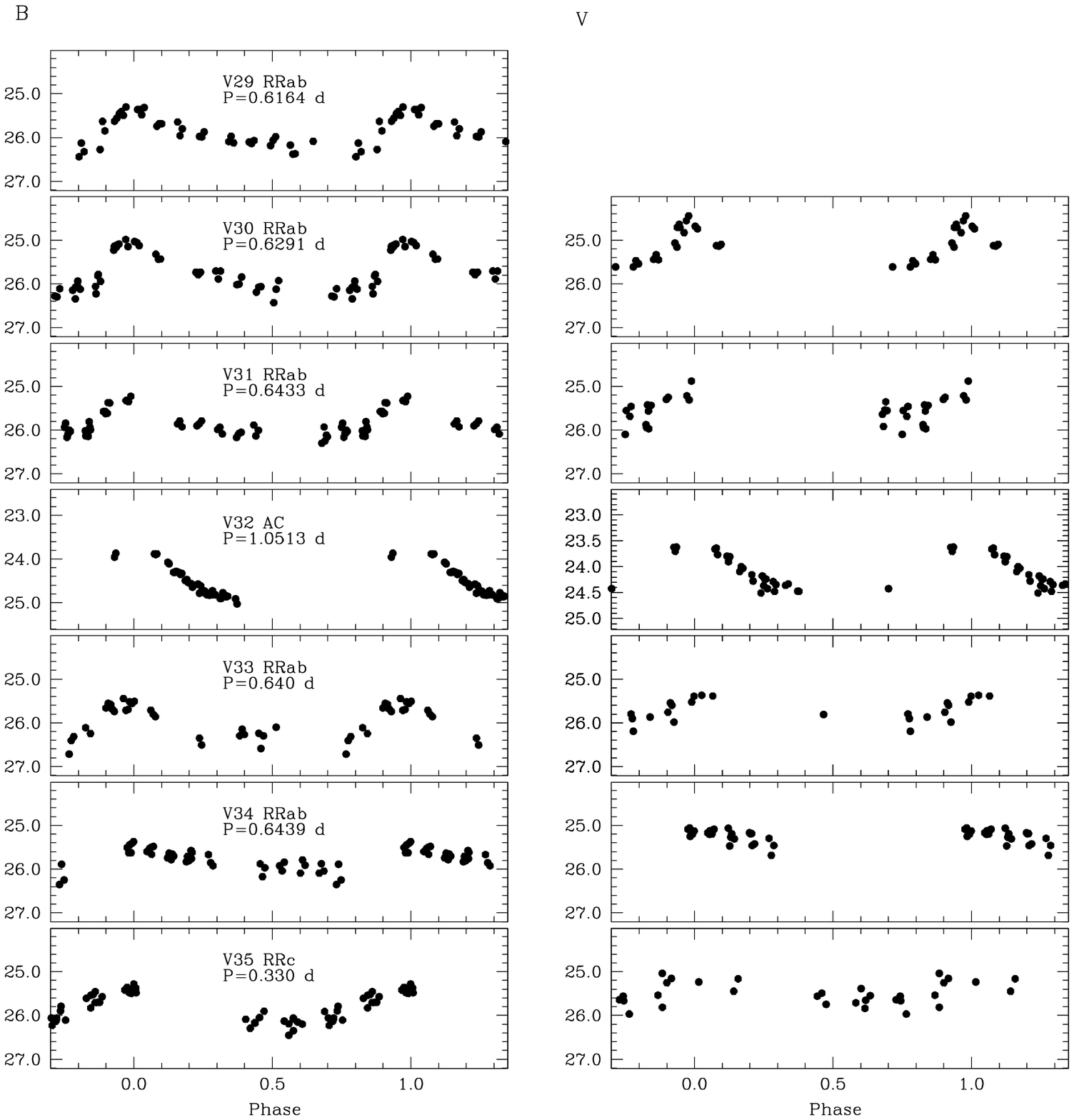}
\caption{ continued --}
\end{figure*}
\begin{figure*}
\centering
\figurenum{1}
\includegraphics[trim=0.001mm 6cm 0.001mm 0.001cm, keepaspectratio=true, width=14cm,height=16cm]{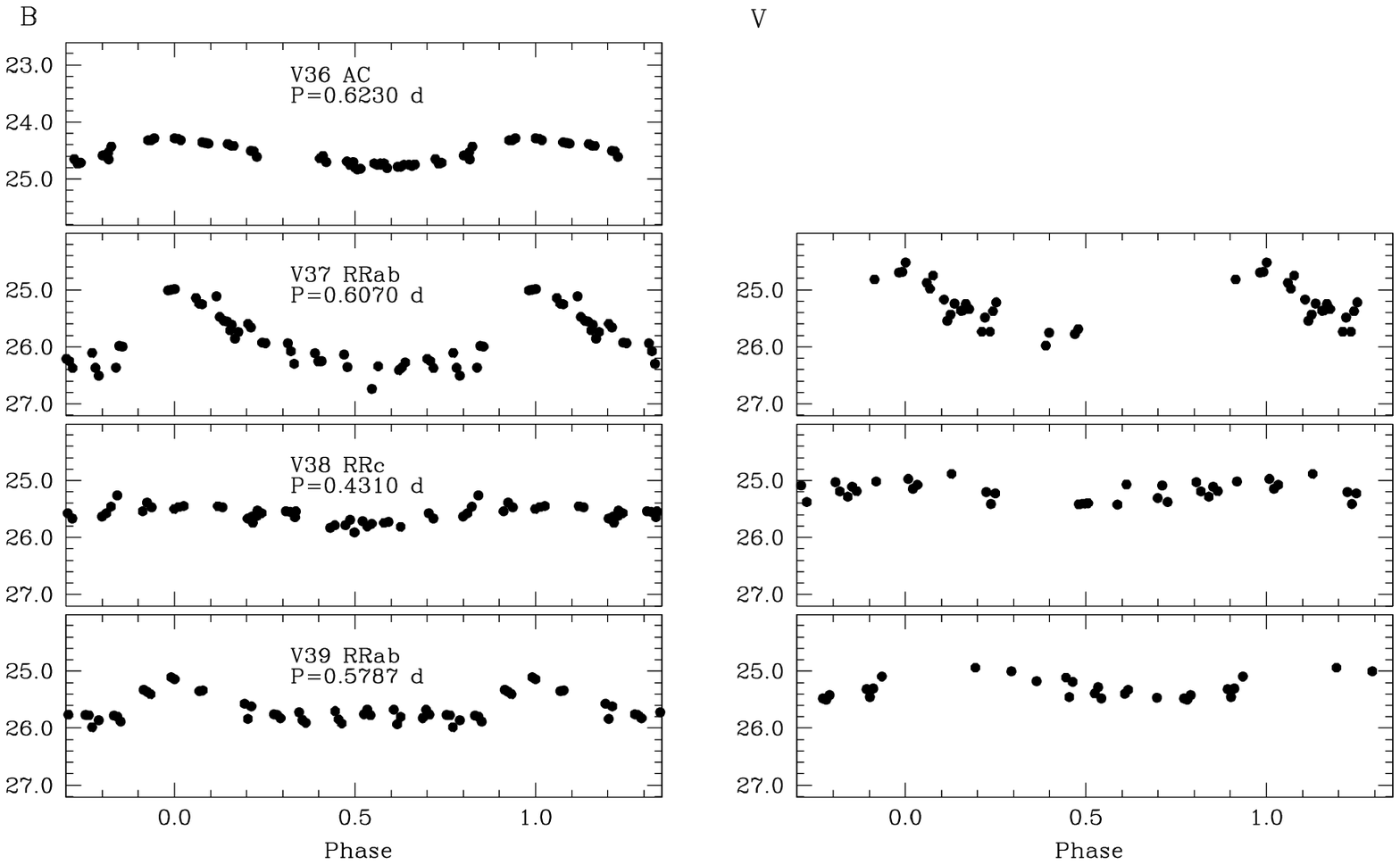}
\caption{ continued --}
\end{figure*}
The time-series photometry of  the variable stars is provided in Table~\ref{t:pho}, which is published in its entirety in the electronic edition of the Journal.

\begin{table*}
\begin{center}
\caption[]{Identification and properties of the variable stars identified in And~XIX }
\footnotesize
\label{t:1}
\begin{tabular}{l c c c c c c c c c c c}
\hline
\hline
\noalign{\smallskip}
 Name & $\alpha$ &$\delta$ & Type & P    & Epoch (max) & $\langle$ B $\rangle$ & $\langle$ V $\rangle$ & A$_{B}$ & A$_{V}$\\
      &	(2000)   & (2000)  &	  &(days)& JD (2400000)& (mag)                 & (mag)                 &  (mag)  &  (mag)  \\
 	    \noalign{\smallskip}
	    \hline
	    \noalign{\smallskip}
%V1  & 00:19:29.6  &+35:02:00.3  & Bin    & & & & & & \\ %12583   
V1  & 00:19:36.1  &+35:02:25.6  & RRab   & 0.558 &55533.592 & 25.81& 25.32&1.49 & 1.17\\%13115	
V2  & 00:19:36.0  &+35:04:17.5  & RRab   & 0.5856& 55477.717& 25.86& 25.46&0.66 &0.53 \\ %15608  
V3  & 00:19:32.4  &+35:00:35.3  & AC     & 1.306& 55477.700& 24.29& 23.82 &1.33 & 1.06\\ %10763
V4$^a$  & 00:19:38.9 & +35:04:47.2  & RRab & 0.649  &  55532.600 &25.89 & $-$     & 0.68  &   $-$    \\%16272  
V5  & 00:19:34.7  &+35:05:30.4  & AC     & 1.043& 55476.140& 24.46&23.94 & 0.66& 0.61\\ %17166   
V6  & 00:19:21.6  &+35:01:01.9  & RRc    &0.3887 &55532.730 & 25.79 &25.31 & 0.53&0.42 \\ %11321 
V7$^a$  & 00:19:44.8 & +35:02:40.5  & RRc    & 0.3905&55532.682 & 25.96 &  $-$      & 0.35&   $-$    \\%13438   
V8  & 00:19:18.3  &+35:02:05.7  & AC     & 0.5873& 55533.583& 24.43& 23.99&0.87 & 0.69 \\ %12704  
V9  & 00:19:26.3  &+34:59:21.7  & RRab   & 0.6159 & 55533.702& 25.77& 25.24& 0.75 & 0.60\\ %9183	
V10  & 00:19:41.8  &+35:05:20.8  & RRab   & 0.6054 &55531.635 &25.77 & 25.27& 1.10&0.87 \\ %16987  
V11 & 00:19:19.1  &+34:59:38.8  & RRab   &  0.6194&55533.592  &  25.83&25.32 &1.16 & 0.92\\ %9555	
V12$^a$  & 00:19:44.1 & +35:05:53.5  & RRab & 0.689   & 55532.700  &25.62  & $-$       &0.67     &    $-$     \\%17640
V13 & 00:19:37.9  &+35:07:22.3  & RRab   & 0.5860 &55477.735 & 25.86& 25.42  &  0.98& 0.78\\ %19430  
V14   & 00:19:23.1  &+34:57:08.5  & RRc     & 0.4010& 55531.653& 25.69 & 25.26 & 0.56& 0.45 \\%6392
V15 & 00:19:23.8  &+34:56:47.7  & RRab   & 0.5851& 55532.790& 25.80& 25.27&0.77 & 0.61\\ %5974	
V16 & 00:19:58.1  &+35:02:59.6  & RRc    & 0.3748& 55531.795&25.74 & 25.35& 0.64& 0.51\\ %25953  
V17 & 00:19:58.4  &+35:02:57.6  & RRab   & 0.6105& 55533.695& 25.83&25.42 & 1.06& 0.84\\ %25871  
V18 & 00:19:04.7  &+35:02:58.3  & RRab   & 0.6167& 55477.723& 25.80  & 25.35&1.17 & 0.93\\ %18775  
V19 & 00:19:55.0  &+34:58:25.6  & AC     & 1.2150& 55533.830&24.26 & 23.87& 1.54& 1.22\\ %15628  
V20 & 00:19:03.6  &+35:00:32.5  & AC     & 1.146 & 55480.720& 24.43& 23.86&0.80 & 0.63\\ %14818  
V21 & 00:19:29.6  &+34:55:12.2  & RRab   &0.6052 &55531.686&25.63 &  25.27&0.89 & 0.71\\ %3984	
V22 & 00:19:55.5  &+35:07:30.1  & RRc   & 0.4090& 55531.710&  25.76 &25.33 &0.49 &0.45 \\ %5002025
V23 & 00:19:29.8  &+34:54:39.8  & RRc   & 0.4076 & 55477.718& 25.69 & 25.26& 0.57&0.38\\%3297
V24 & 00:19:30.5  &+34:54:29.2  & AC     & 0.5839& 55533.785& 24.90& 24.51& 1.28&1.02 \\ %3080	
V25 & 00:20:06.6  &+35:03:54.8  & RRab    & 0.643& 55477.717 &  25.73& 25.36& 1.02& 0.81\\ %28092  
V26$^a$ & 00:19:33.3 & +35:11:27.7  & RRab & 0.575   & 55532.767  &  25.75   &  $-$     & 1.02&  $-$       \\%11807
V27 & 00:19:55.8  &+34:56:05.9  & RRab   &0.6653  & 55532.735&25.57 & 25.17& 1.06 & 0.85\\ %10245  
V28 & 00:19:47.1  &+35:10:59.9  & RRab   & 0.6423& 55531.742& 25.88& 25.34& 1.07& 0.85\\ %10099  
V29$^a$ &  00:20:10.9 & +35:04:53.5  & RRab &  0.6164 &  55531.780 &  25.74  &  $-$      & 0.81&  $-$     \\%30308
V30 & 00:18:51.7  &+35:04:34.1  & RRab   &0.6291 & 55477.106  & 25.73& 25.25&1.21&0.96 \\ %21566  
V31 & 00:20:10.4  &+35:07:30.4  & RRab   & 0.6433& 55533.793&25.82 &  25.42& 0.85&0.69 \\ %36270  
V32 & 00:18:58.6  &+34:55:38.4  & AC     & 1.0513& 55477.790&  24.47& 24.10& 1.15& 0.92\\ %6918	
V33 & 00:18:59.7  &+35:10:21.1  & RRab   & 0.640 &  55533.733&26.15 & 25.65& 1.11&0.88 \\ %7873	
V34 & 00:20:16.9  &+35:02:24.0  & RRab   &  0.6439& 55531.670& 25.83& 25.35&0.66 & 0.52\\ %24628  
%V31 & 00:19:55.1  &+35:12:31.4  & RRab   & & & & & & \\ %15703  
V35 & 00:19:26.9  &+35:14:24.9  & RRc    & 0.330& 55532.790&25.84 &25.44 &0.77 &0.62 \\ %22433  
%V33 & 00:20:17.2  &+34:57:53.8  & sx Phe?& & & & & & \\ %14377  
V36$^a$ & 00:20:03.2  &+35:12:33.9  & AC     & 0.6230& 55532.650&  24.55 &$-$  & 0.48&  $-$ \\ %15873  
V37 & 00:18:40.8  &+35:05:14.3  & RRab   & 0.6070& 55531.669&25.82 &25.30 & 1.50& 1.19\\ %22765  
V38 & 00:18:59.3  &+35:12:59.7  & RRc   & 0.4310&55533.810 & 25.59& 25.21&0.33 &0.32 \\ %17450  
V39 & 00:20:19.7  &+34:52:48.8  & RRab   &0.5787 &55532.750 & 25.64 &25.31 &0.75 &0.60 \\ %3233	
%V  & 00:19:44.1 & +35:05:53.5  & RRab & 0.689   & 55532.700  &onlyB  &        &     &       \\%17640
%V  & 00:19:33.3 & +35:11:27.7  & RRab & 0.575   & 55532.767  &onlyB  &        &      &       \\%11807
%V  & 00:19:38.9 & +35:04:47.2  & RRab & 0.649  &  55532.600  &onlyB  &        &       &     \\%16272
%V  & 00:19:44.8 & +35:02:40.5  & RRc  &  0.3905 &  55532.682 & onlyB &        &         &     \\%13438  
%V &  00:20:10.9 & +35:04:53.5  & RRab &  0.6164 &  55531.780 & onlyB &       &          &     \\%30308
\hline
\end{tabular}

Notes:\\
 $^a$ Only $B$ light curves are available for these stars, the $\langle$B$\rangle$  values were obtained by adding to the instrumental 
 $\langle$b$\rangle$  values the zero point of the $B$ calibration equation.\\

\normalsize
\end{center}
\end{table*}

\begin{table*}
\begin{center}
\caption{$B,V$ photometry of the variable stars detected in And~XIX. This
Table is published in its entirety in the electronic edition of the
Journal.
A portion is shown here
for guidance regarding its form and content.}
\label{t:pho}
\vspace{0.5 cm}
%\small
\begin{tabular}{cccccc}
\hline
\hline
%\multicolumn{1}{c}{}& \multicolumn{1}{c}{}&
\multicolumn{6}{c}{And~XIX - Star V1 - {\rm RRab}} \\
%\begin{tabular}{cccc}
%\hline
%\noalign{\smallskip}
%& & Star V18 - {\rm RRab}&  \\
\hline
{\rm HJD}  & B & {\rm $\sigma_B$} & {\rm HJD } & V &{\rm
$\sigma_V$} \\
($-$2455000) & {\rm (mag)} & {\rm (mag)}  & ($-$2455000) & {\rm (mag)} &
{\rm (mag)}  \\
%            \noalign{\smallskip}
\hline
%\noalign{\smallskip}
477.72208  &   26.03  &  0.30  & 531.66930 & 25.69 & 0.23 \\
531.65805  &   26.51  &  0.30  & 531.75167 & 25.85 & 0.18 \\
531.66338  &   26.52  &  0.24  & 531.79774 & 25.69 & 0.19 \\
531.66935  &   26.39  &  0.27  & 531.80311 & 25.65 & 0.27 \\
531.70489  &   26.56  &  0.32  & 533.58629 & 24.76 & 0.14 \\
531.71021  &   26.47  &  0.35  & 533.59162 & 24.81 & 0.13 \\
531.71570  &   26.34  &  0.24  & 533.59698 & 24.52 & 0.15 \\
531.75178  &   26.17  &  0.18  & 533.63227 & 25.02 & 0.08 \\
531.75713  &   26.24  &  0.18  & 533.63763 & 24.93 & 0.10 \\
531.76244  &   26.25  &  0.16  & 533.64302 & 24.90 & 0.07 \\

\hline

\end{tabular}

\end{center}

\end{table*}

\begin{figure*}[!t]
\centering
\includegraphics[scale=0.9]{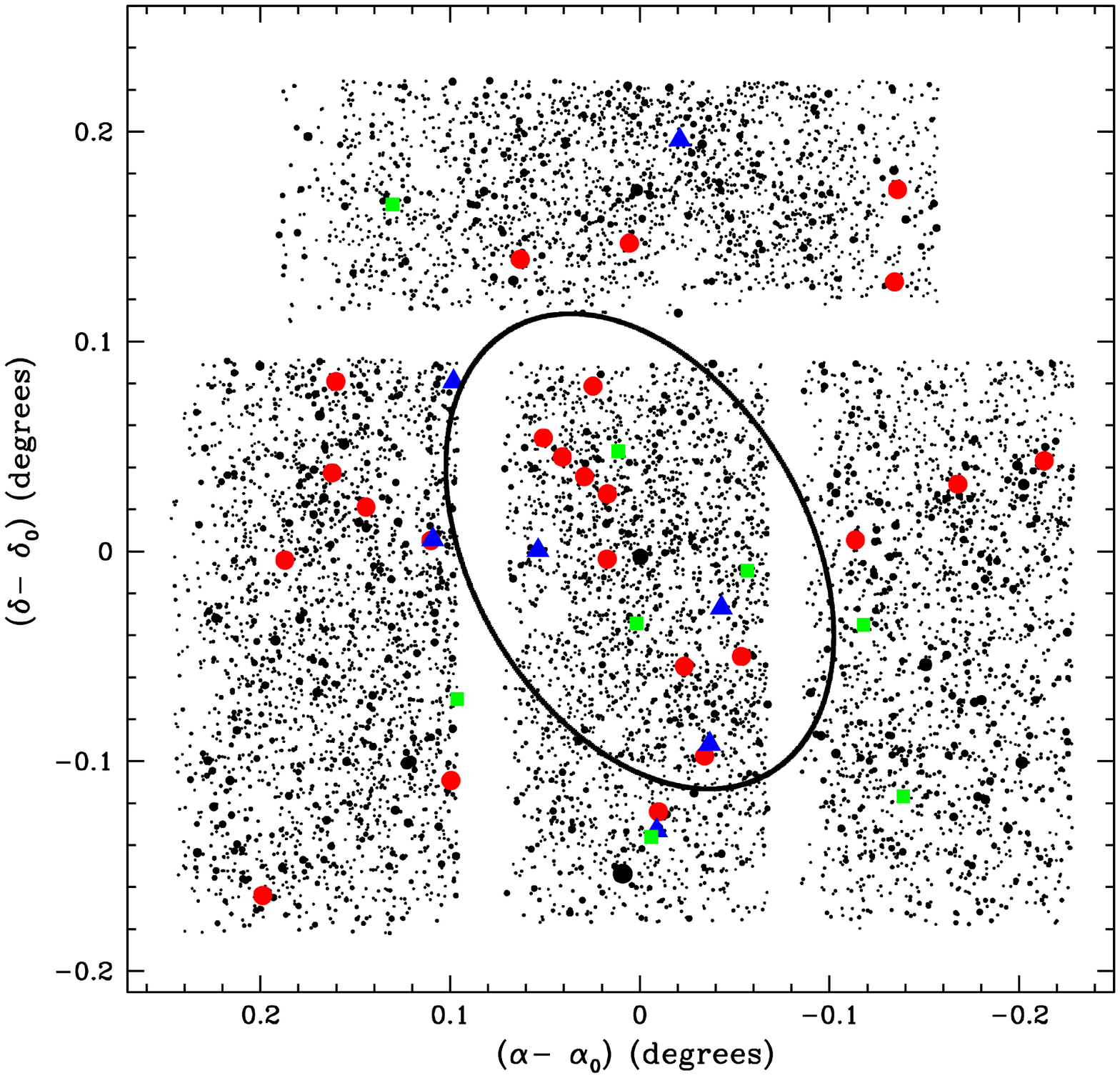}
\caption[]{Spatial distribution of  the variable stars identified in the FoV covered by our LBC observations of And~XIX.
Red circles and blue triangles mark RRab and RRc stars, respectively.  Green squares are ACs. Black dots are non-variable stars: 
they are plotted with symbol sizes inversely proportional to their magnitude. The black  ellipse is drawn by convolving And~XIX half-light radius  with the galaxy  ellipticity and position-angle
following  \citet{mac08} .}
\label{fig:map}
\end{figure*}

Figure~\ref{fig:map} shows the position of the variable stars in the  FoV covered by our LBC observations. In the figure  
black dots mark non-variable stars,  red circles and blue triangles 
mark  RRab and RRc stars, respectively, while the ACs are shown by green squares. 
A black  ellipse drawn by convolving And~XIX half-light radius  with the galaxy  ellipticity and position-angle, as  published by \citet{mac08}, 
delimitates 
the region containing the bulk of And~XIX's stars. In the following we refer to this area as the region within And~XIX half-light radius.

\section{PERIOD-AMPLITUDE DIAGRAM AND OOSTERHOFF CLASSIFICATION}\label{sec:oo}

\begin{figure*}[!t]
\centering
\includegraphics[keepaspectratio=true, scale=0.62]{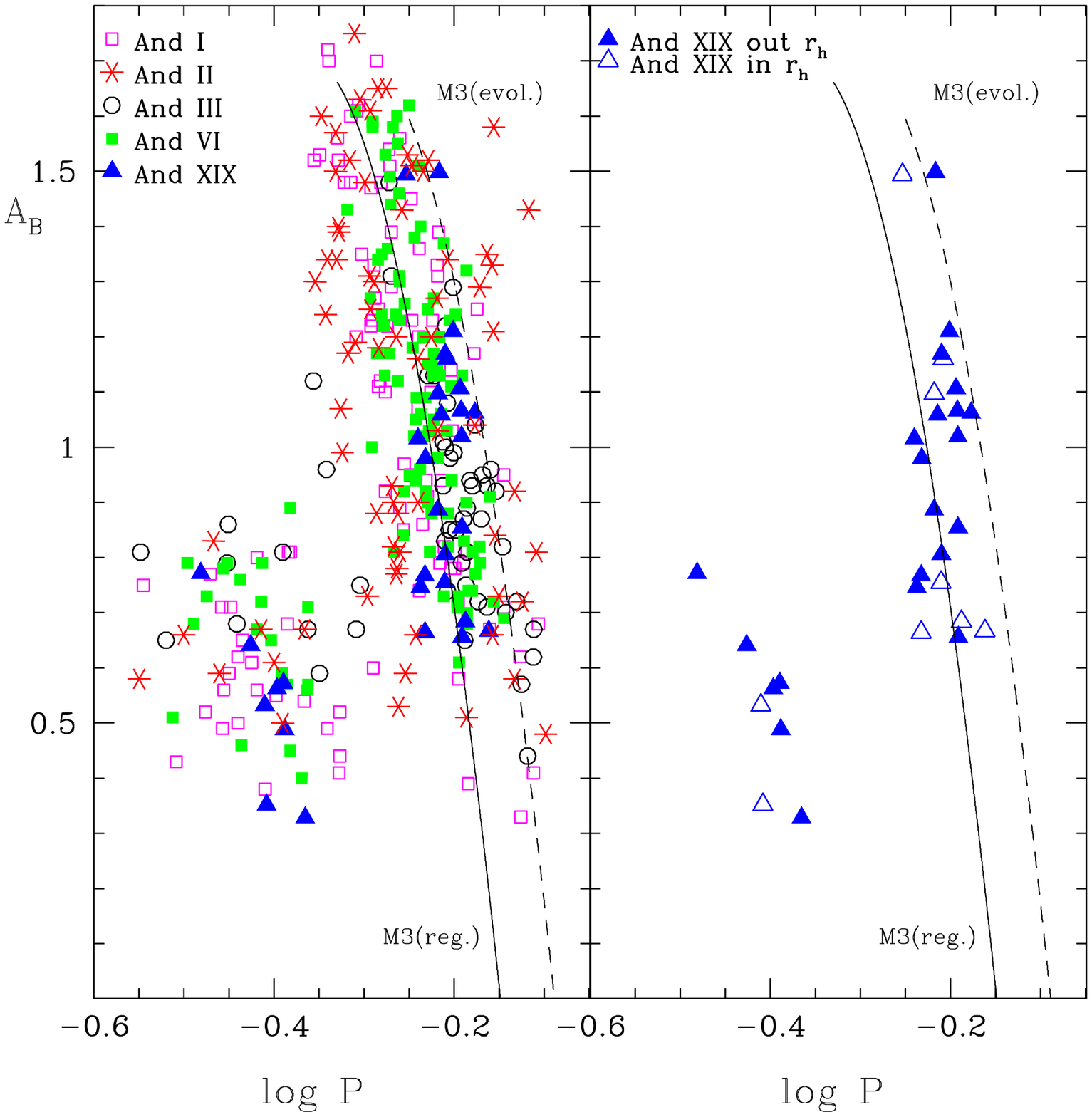}
\caption[]{ $Right$: Period-amplitude diagram in the $B$-band for the RR Lyrae stars identified in the present  study.   Open symbols mark variables falling inside the ellipse 
shown in Fig.~\ref{fig:map}. 
%defined  by
%convolving And~XIX half-light radius with the galaxy ellipticity.
Also shown, for comparison, are the loci defined by the 
bona-fide regular (solid curve) and well-evolved (dashed curve) fundamental-mode RR Lyrae stars in M3, from Cacciari et al. (2005); 
$Left $: comparison of And~XIX  variables with the RR Lyrae stars stars identified in  other four   M31 dwarf satellites (namely, And~I, And~II, And~III and And~VI).  %Data for And~I and And~III  are from \citep{pri05}, for And~II  from \citep{pri04} and for And~VI from \citep{pri02}. The lines  defining the Oo-I and Oo-II classes
%are from \citet{cac05}. 
}
\label{fig:bay}
\end{figure*}

The right panel of Figure~\ref{fig:bay} shows the $B$-band period-amplitude diagram (also known as Bailey diagram, \citealt{Bai1902}) of the  
RR Lyrae stars identified in this study.  Open symbols mark variable stars falling inside the  ellipse shown in Figure~\ref{fig:map}. Variables inside and outside the galaxy
half-light radius do not show differences in this plot.
%
%defined  by convolving And~XIX half-light radius with the galaxy ellipticity, according 
%to \citep{mac08}. 
Also shown in the figure %~\ref{fig:bay}
 are the loci defined, respectively, by the 
bona-fide regular (solid curves) and the well-evolved (dashed curves) fundamental-mode RR Lyrae stars in the Galactic  globular cluster (GC) M3, from Cacciari et al. (2005). M3 has Oo~I  properties 
and its regular RRab stars  define very well the 
locus occupied by the MW Oo~I GCs, whereas the evolved RRab stars mimic the locus defined by the MW Oo~II GCs.
All the RR Lyrae stars identified in this study, independently of being located inside or outside And~XIX's half-light radius,  fall on  either the Oo~I locus, or between the two lines defined by the M3 variables \citep{cac05}. 
The left panel of Figure~\ref{fig:bay} shows the comparison of And~XIX  variables with the 
period-amplitude diagram of the  RR Lyrae stars detected in other four   M31 dwarf satellites (namely, And~I, And~II, And~III and And~VI).
Data for the RR Lyrae stars in And~I and And~III are from \cite{pri05}, for And~II  from \cite{pri04}, and for And~VI  from \cite{pri02}.
 The variables appear to concentrate mainly towards the Oo-I locus or in the region between Oo~I and Oo-Int lines, 
thus suggesting that there is paucity of Oo-II satellites around M31. 

The average period of the RRab stars inside the  ellipse in Figure~\ref{fig:map} is $ \langle {\rm Pab} \rangle =$0.610 d (average on 7 stars,  $\sigma=0.03$ d) and becomes 
$\langle {\rm Pab} \rangle =0.62$ d  (average on 23 stars,  $\sigma=0.03$ d)   if we average over all the  RRab stars identified in the present study. This along with the position on the period-amplitude diagram suggest that 
all the RR Lyrae stars we have identified likely belong to And~XIX. 
%23 RRab stars in And~XIX 
% is $\langle {\rm Pab} \rangle =0.62$ d  (average on 23 stars,  $\sigma=0.03$ d)  and becomes  
%$ \langle {\rm Pab} \rangle =$0.610 d (average on 7 stars,  $\sigma=0.03$ d)
% if we consider only RRab stars inside the  ellipse % obtained by convolving And~XIX's half-light radius with the galaxy ellipticity (see 
%  in Figure~\ref{fig:map}. 
The ratio of number of RRc over   total number of RR Lyrae stars is $f_c$=0.26. Based on these 
evidences  we conclude that %(see Sect.~\ref{sec:intro}) 
And~XIX has Oo-Int properties.

\section{DISTANCE}\label{sec:dist}

We measured the distance to And~XIX  using its RR Lyrae  stars. 
The average $V$ magnitude of the RR Lyrae stars is $ \langle {\rm V(RR)} \rangle =25.34$ mag ($\sigma=0.10$ mag, average over 26 stars).
If we consider only RR Lyrae stars inside the galaxy half-light radius 
(ellipse in Fig.~\ref{fig:map}; namely, the 7 RRab stars:  V1, V2, V4, V9, V10, V11,  V12; and the two RRc stars 
V6, V7) the average becomes  $ \langle {\rm V(RR)} \rangle =25.32$ mag ($\sigma=0.07$ mag, average over 9 stars). 
The difference between the two average values
is negligible, thus further confirming that all the RR Lyrae stars we have identified likely belong to And~XIX.
In the following we will use the  average over all the RR Lyrae stars as more representative of the whole  galaxy. 
The $ \langle {\rm V(RR)} \rangle $ value was de-reddened using a standard extinction law (A$_{\rm V}$=3.1$\times$E(B-V)) and as a first 
approach  the reddening value   E(B-V)=0.066$\pm$0.026 mag provided by  \citet{sch98} maps.
% To derive an estimate of the distance of And~XIX
 We then assumed an absolute magnitude  of M$_{\rm V}=0.54\pm0.09$ mag 
for RR Lyrae stars at [Fe/H] = $-$1.5 dex (\citealt{cle03};   this is consistent with a  distance modulus for the 
Large Magellanic Cloud, LMC,  of 18.52$\pm 0.09$ mag)   and  corrected for the different metal abundance using 
the relation $\frac{\Delta {\rm M_V}}{\Delta {\rm[Fe/H]}}=0.214\pm0.047$ mag/dex by \citet{cle03} and \citet{gra04}. We adopted  
for And~XIX the metallicity [Fe/H]=$-1.8 \pm 0.3$ dex derived spectroscopically by \citet{col13}. 
The distance modulus of And~XIX derived under the above assumptions is:  (m-M)$_0$=24.66$\pm$0.17 mag.
An independent reddening estimate can be obtained from the RR Lyrae stars using \citet{pier02}'s 
method which is based on the relation between intrinsic $(B-V)_0$ color, period,  metallicity and $B$-amplitude of the RRab stars. % we derived an independent estimate of the reddening for And~XIX.
Using the 19 RRab stars for which we  have  photometry in both the $B$ and $V$ bands, we obtain
E(B-V)=$0.11\pm0.06$ mag. %Using this last value  we estimate the distance following the same procedure 
%as before. 
The distance modulus derived with this new reddening is: (m-M)$_0$=$24.52\pm0.23$ mag. Both our distance 
moduli  place And~XIX 
almost at the same distance of M31 and both are in  good agreement, within the errors, 
with the value of (m-M)$_0=24.57^{+0.08}_{-0.43}$ mag found by \cite{con12}.  Our  estimates are instead 
smaller than  \citet{mac08} modulus  for And~XIX, but  still consistent with their value within 1$\sigma$.

\section{THE CMD}

\begin{figure*}[!t]
\centering
\includegraphics[width=\textwidth,height=\textheight,keepaspectratio]{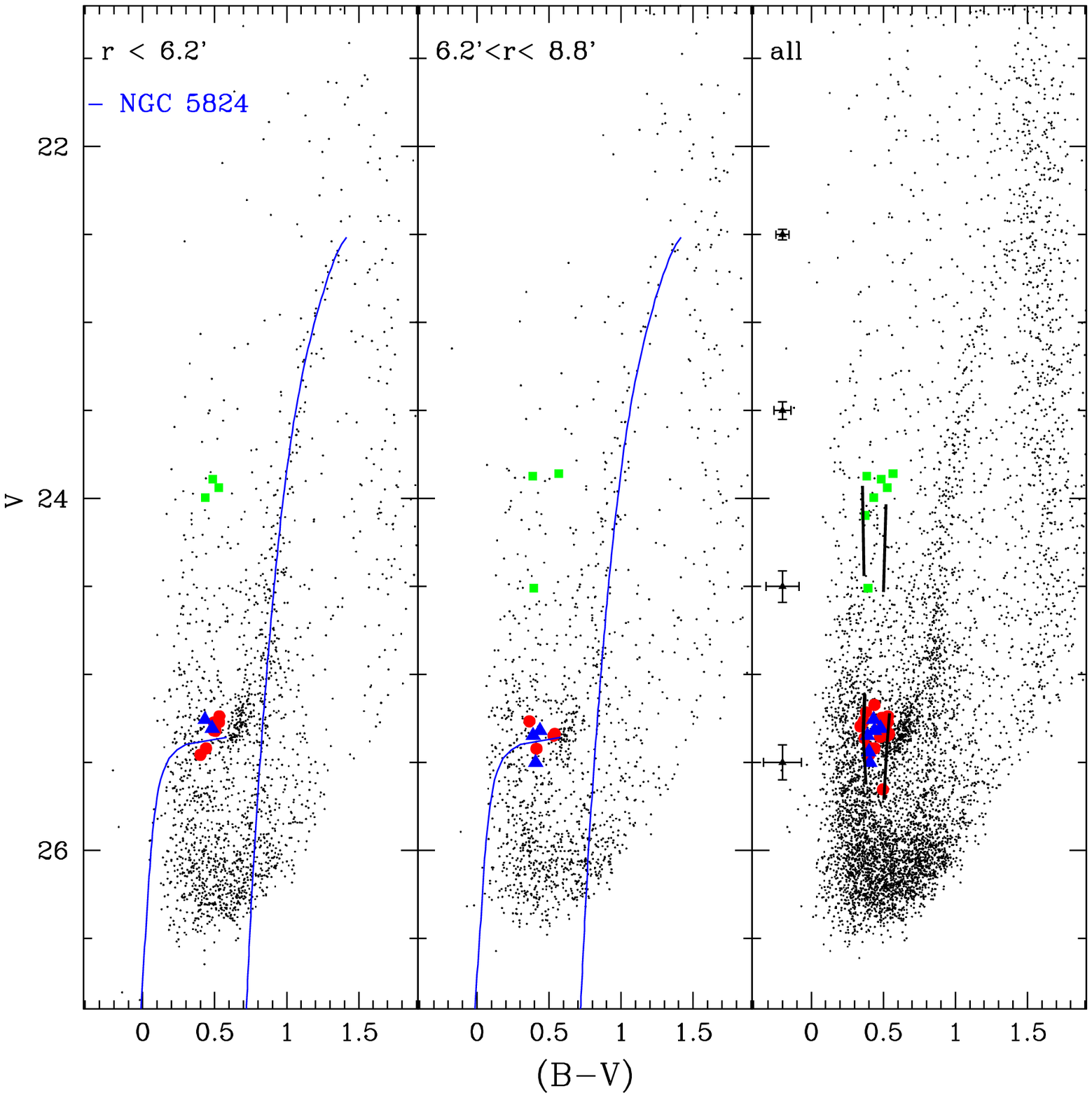}
\caption[]{
CMDs obtained selecting objects in different regions of the FoV. Only the 
%To avoid contamination from background  galaxies and peculiar objects,
% the object plotted in the CMD 
%were selected from 
%we selected our photometric catalog using the $\chi$ and Sharpness parameters provided by \texttt{ALLFRAME}. 
sources for which  $-0.3 \le$ Sharpness $\le 0.3$ and $\chi < 1.0 $ for magnitudes fainter than $V$ = 22.0 mag, and $\chi$  $<$ 1.5 for magnitudes 
brighter than $V \sim$ 22.0 mag, are displayed in the three panels of the figure, in order to reduce contamination from background  galaxies and peculiar objects. 
 Red circles are RRab stars, blue triangles are RRc stars, and green squares are ACs. 
 $Left $: CMD of only  stars inside an ellipse drawn with the galaxy half-light radius (see Fig.~\ref{fig:map}).
% selected from  our photometric catalog using the  $\chi$ and Sharpness parameters (see Sect. 6). 
%In this panel
% only stars inside the half-light radius are plotted. 
 Shown in blue are the ridgelines of the Galactic globular cluster NGC5824 ([Fe/H]=$-$1.94 dex,  \citealt{carretta09}); 
$Middle$: same as in the left panel, but for stars inside  an elliptical ring with internal radius 6.2$'$ and external radius 8.8$'$, which encloses the same area as in the left panel; 
 $Right$: CMD of stars in the whole LBC FoV. Black solid lines show the boundaries of the IS for RR Lyrae stars and ACs with $Z$=0.0002, from \citet{mar04}.}
\label{fig:dia}
\end{figure*}

Figure~\ref{fig:dia} shows the CMDs obtained in the present study selecting objects in different regions of the FoV.

To avoid contamination from background  galaxies and peculiar objects,
% the object plotted in the CMD 
%were selected from 
we selected our photometric catalog using the $\chi$ and Sharpness parameters provided by \texttt{ALLFRAME}. 
We only retained sources for which  $-0.3 \le$ Sharpness $\le 0.3$ and with $\chi < 1.0 $ for magnitudes fainter than $V$ = 22.0 mag, and $\chi$  $<$ 1.5 for magnitudes 
brighter than $V=$ 22.0 mag. These selections led to a total of $\sim 9000$ stars, plotted as dots in Fig.~\ref{fig:dia}.
% were selected as to retain only sources with $\chi$  $<$ 1.5 and $-0.3 \le$ Sharpness $\le 0.3$. 
The variable stars are 
plotted in Fig.~\ref{fig:dia} according to their intensity-average magnitudes and colors and using red circles for the  RRab stars, 
blue triangles for the  RRc stars,  and green squares for the ACs. Only 26 RR Lyrae stars and 7 ACs could be plotted, as we lack $V$ photometry for 
5 RR Lyrae stars and 1 AC. 
The right  panel of Figure~\ref{fig:dia} shows the CMD of  the whole FoV of  the LBC observations;  the 1$\sigma$ error bars, as derived from %extensive
 artificial 
star tests conducted on real images, are also drawn. The left panel shows only  
stars inside the ellipse drawn with  the galaxy half-light radius (r$_h$=6.2$'$, \citealt{mac08}; see Fig.~\ref{fig:map}),  
and the central panel  shows the CMD of  stars inside an elliptical ring with internal radius 6.2$'$ and external radius 8.8$'$, 
which encloses the same area as in the left panel. %The radius is defined like $r^2 = {{x}^{2} + {y}^{2}/{(1-e)^{2}}}$ where $e$ is  the ellipticity   of the galaxy 
% taken  from \citet{mac08}. 
The most prominent features of the And~XIX CMD are:

\begin{itemize}
\item a  RGB, between $B-V=0.6-1.5$ mag, extending upwards to
$V \approx$ 22-22.5 mag;

\item a red HB with colors $0.6<B-V<0.8$ mag;

\item  a % blue plume, 
 distribution of sources extending nearly vertically at $B-V\sim 0.2-0.4$ mag; the
unusual color of these objects  (see Sect.\ref{spatial}),  %(see, redder than solar isochrones,
 can be explained by unresolved background galaxies; % (see Sect.\ref{spatial});

\item a sequence of stars at $B-V\sim 1.5-1.7$ mag and extending blueward
around $V=22$ mag. This feature is likely composed by foreground field stars (also see Section~\ref{spatial}).

\end{itemize}
In the left and center panels of Fig.~\ref{fig:dia} we have plotted in blue the ridge lines of the Galactic globular cluster  
NGC 5824 from   \citet{pio02},  corrected to the distance and  reddening of And~XIX 
we derived from the RR Lyrae stars (see Sect. \ref{sec:dist}). 
We assumed for NGC 5824 a distance modulus of (m-M)$_0$=17.54  mag and a reddening  of E(B-V)=0.13 mag  from \citet{har10} catalog (2010 edition).
NGC 5824 has a metallicity   [Fe/H]=$-$1.94 $\pm$ 0.14 dex \citep{carretta09} that is very similar to the metal abundance of And~XIX estimated by \citet{col13}.
The RGB of  NGC 5824 matches very well And~XIX's  RGB thus confirming both the similar metallicity and the higher reddening value inferred from the RR Lyrae stars.
% and the ridgeline of NGC 5824 and the presence of RR Lyrae stars indicate
%that  the galaxy has an old population with an age $>$ 10 Gyrs. 
The IS boundaries for RR Lyrae stars and ACs with $Z$= 0.0002 from \citet{mar04} are overplotted to the CMD  in the right panel of Fig.~\ref{fig:dia}.
The  variables we have classified as RR Lyrae stars fall well  inside the boundaries of the RR Lyrae IS, 
confirming they are bona-fide RR Lyrae stars.   
Similarly,  the variables above the HB appear to be confined in the region of the CMD where 
 ACs  are usually found (see following section). %Sect.~\ref{sec:ACS}).

\section{ANOMALOUS CEPHEIDS}\label{sec:ACS}

\begin{figure*}[!t]
\centering
\includegraphics[width=\textwidth,height=\textheight,keepaspectratio]{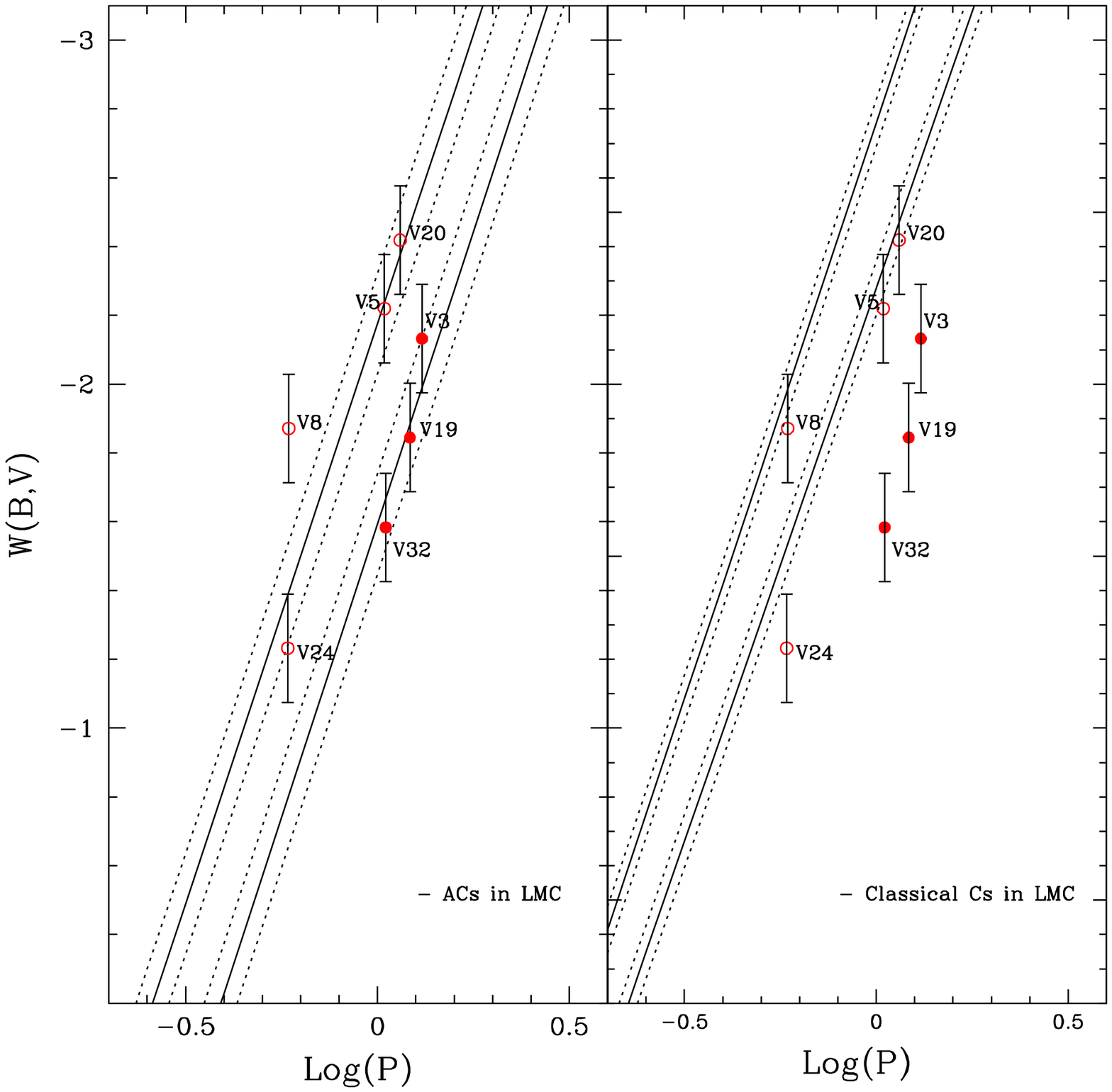}
\caption[]{{\it Left panel}: $PW$ relations for And~XIX's variables brighter than the HB. Solid lines are the fundamental-mode
  (lower line) and the first-overtone (upper line)
$PW$ relations for ACs by \citet{ripe13} \citep[converted to $B$, $V$ bands using the relations in ][]{mar04},  
with their related 1$\sigma$ uncertainties. {\it Right panel}: same as left panel, but with the 
$PW$ relations for Classical Cepheids by \citet{sos08b}}
\label{fig:plac}
\end{figure*}

In And~XIX we have identified 8 variable stars from about 1 to 1.5 mag brighter
than the average $B$ magnitude of the HB.  These stars are found to  fall inside the  boundaries of the IS for ACs from \citet{mar04} (see Fig.~\ref{fig:dia}), thus 
providing support to their classification as ACs. In order to investigate further their nature we have also compared them with the
$PL$ relations for ACs.
ACs  %are also good distance indicators,  as they  
follow
 a Period-Luminosity ($PL$) relation which differs from both the 
Classical Cepheids and  the type II Cepheids $PL$ relationships 
\citep[see Fig.~1 of ][]{sos08a}.
%The ACs  have higher mass and lower metallicity when compared to P2C, and for this reason at a given
%luminosity they have shorter period than P2Cs. 
Unfortunately, the $PL$ relation has the disadvantage of being reddening dependent and in some cases the scatter around
the mean value can be very high. Narrower relations are found introducing a color term in the $PL$ relation and in particular  
 the Wesenheit function \citep{van75,mad82}  includes a color term whose coefficient is equal to the ratio between 
total-to-selective extinction in a filter pair. 
In such a way the $PW$  relation  is  reddening free by definition. 
The Wesenheit index in our case is  W($B,V$)= M$_V -$ 3.1$\times$($B-V$), where  M$_V$ is the $V$ magnitude 
corrected for the distance.
We have  $B$ and $V$ magnitudes for  7 of the variables above the HB. Their $\langle V \rangle $ magnitudes were corrected using the distance
modulus (m-M)$_0$=24.52 mag  derived from the RR Lyrae stars and used to derive the corresponding Wesenheit indices.
The position of these 7 variables in  the  $PW$ plane is shown in the left panel of Figure~\ref{fig:plac}, where we also plot, as solid lines,  
the $PW$  relations for ACs recently derived by  \citet{ripe13} using 
 25 ACs in the LMC\footnote{\citet{ripe13}'s relations were derived for the $V$ and $I$ bands, 
and we have  converted them to $B$ and $V$ 
 using Equation 12 of \citet{mar04}.} (LMC). 
Six of And~XIX  bright variables appear to fall well on  the   \citet{ripe13}'s $PW$ 
relations for ACs with stars V3, V19, V32 likely being fundamental-mode pulsators, and  stars V5, V20, V24 
likely pulsating in the first-overtone mode. On the other hand, star V8 appears to be more than 2$\sigma$ off
 the  first-overtone $PW$ relation, hence its classification as AC seems to be less robust.
 To show that these bright variables are mostly ACs, on the right panel of Figure~\ref{fig:plac} 
the $PW$ relation for Classical Cepheids (CCs) in the LMC 
derived by \citet{sos08b} is shown. The $PW$ relation for CCs indeed does not fit very well the 
  bright variables in And~XIX. 

%and the bright variables in  AndXIX is very good, and 
% from this we can also classified 3 fundamental-mode and 4 first-overtone pulsators.
 % To conclude ...
 We will further discuss the nature of the bright variables of And~XIX in Sect.~\ref{sec:recent} where we compare them with theoretical isochrones.
 
\begin{figure*}[!t]
\centering
\includegraphics[width=\textwidth,height=\textheight,keepaspectratio]{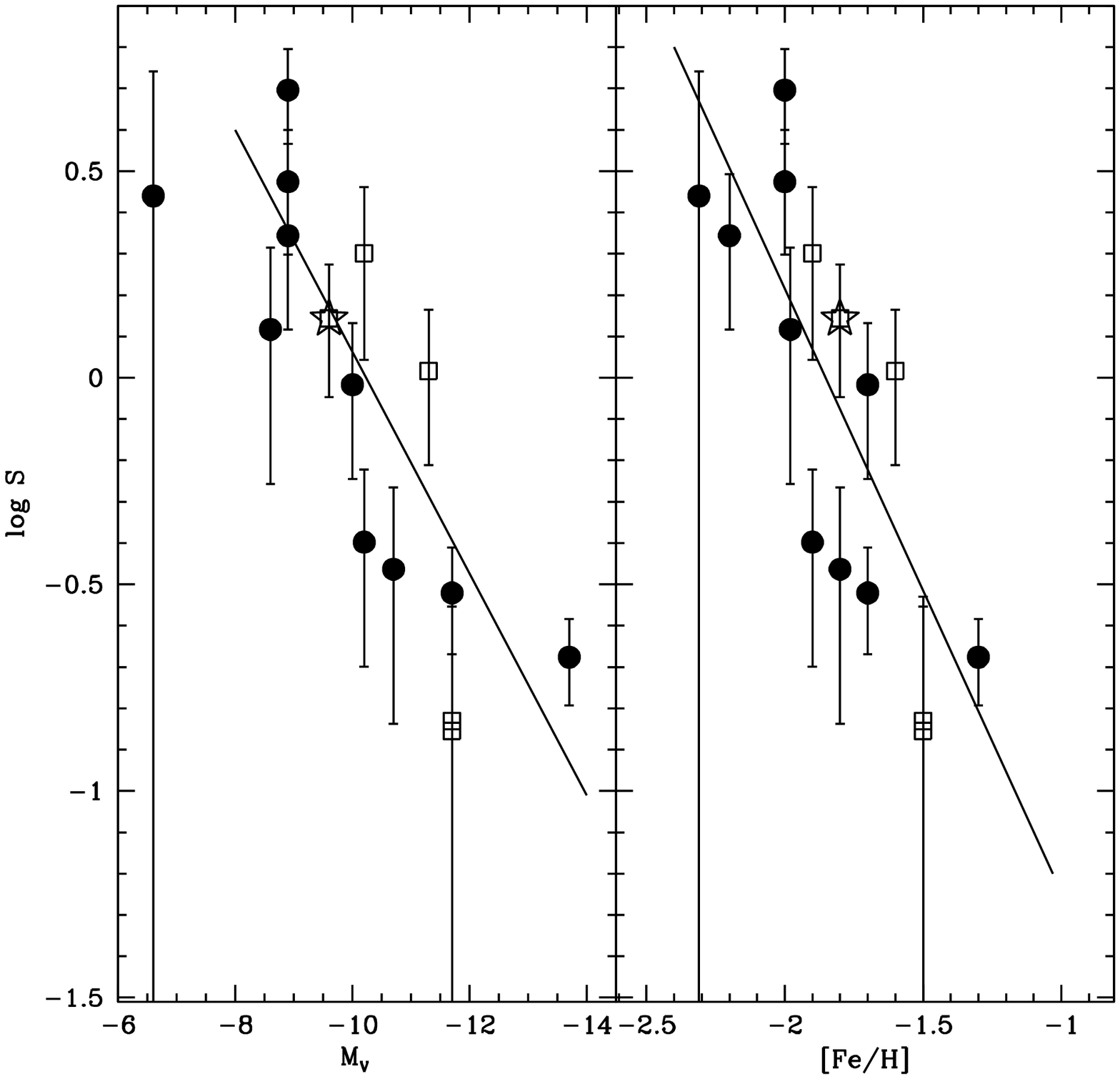}
\caption[]{Specific frequency of ACs in MW (filled circles) and M31 (open squares)  dSphs vs. absolute visual 
 magnitude (${\rm M_V}$, \textit{left panel}),  
and  metallicity ([Fe/H], \textit{right panel}) of the parent galaxy. %The Galactic dSphs are indicated with  filled circles, while the M31 dSphs with  open squares. 
And~XIX is marked by a star symbol.}
\label{fig:ac}
\end{figure*}

 \citet{mat95} found that the specific frequency of ACs
(i.e. the number of ACs per 10$^5$ L$_V$) in the Galactic dSph
galaxies is related to the luminosity and metallicity of the parent
dSph. %More recently
\citet{pri04,pri05} found that this correlation also holds 
for the M31's satellites And~I, And~II, And~III and And~VI.  On the assumption that the 8 supra-HB variables of And~XIX are ACs, 
in Figure~\ref{fig:ac} we plot 
 the  specific frequency of ACs in And~XIX versus its luminosity (left panel) and metallicity (right panel) 
%\citep[from][]{mac08} 
and compare it to the  one in 
Galactic and M31 dSphs from \citet{pri04}. 
And~XIX follows well the relation traced by the other dSphs. This also indicates 
that  we have detected almost all the  short period
variable stars at $V \sim 24$ mag in And~XIX, as also dictated by the 
 artificial star test that gives a completeness of $\sim80\%$ at this level of magnitude.

%together with a collection of  ACs in different dSph galaxies
%%\citep{pri02,pri04} and Leo T \citep{cle12}. The dotted lines represent the fit to the first overtone and fundamental mode pulsators 
%obtained with the ACs plotted.
%Six of the eight ACs of And~XIX are well placed over the fundamental mode 
%fit line, while one is probably a first overtone pulsator. This result give confidence to our distance and reddening estimates.    

\section{M31 HALO CONTAMINATION}

Although And~XIX is  far ($\sim 120$ kpc, \citealt{con12})  from the M31 center 
the contamination from RR Lyrae stars and ACs belonging to the M31 halo 
may be not negligible. The FoV of the LBC ($\sim 0.15~\rm{deg}^2$)
is such that we were able to fit in just one pointing a large portion of And~XIX ($\simeq~2~\times~r_h$).
Unfortunately, in such a large FoV contaminants are also expected to be present in a large number. As we showed in Sect.~\ref{sec:dist}
And~XIX is almost at the same distance of M31 and for this 
reason the luminosity of the variable stars in And~XIX and in the M31 halo is
similar, thus distinguishing the two samples  on the basis of the average luminosity is not possible. 
 To estimate how many variable stars belonging to the M31 halo
can be  expected to contaminate And~XIX's sample we used the results from the search for variable stars
in 6 fields around  M31 made by  \citet{jef11} using HST/ACS. Two of these fields are in the halo of M31
 at a  distance of about 35 kpc from the  center  in the  south-east direction along the galaxy minor axis. They were found  
to contain 5 and no RR Lyrae stars, respectively, despite they have comparable stellar density. Furthemore, 
none of the two halo fields was found to contain ACs.  
From these results
we can give a rough estimate of how many RR Lyrae stars and ACs  belonging to M31 we expect  to find in the field of And~XIX.
 Assuming a stellar density of the M31 halo of $\propto$ r$^{-3.2}$ \citep{gil12} and 
 after scaling for the different area surveyed by the LBC  and HST/ACS 
we estimated a contamination from the M31 halo by 5 RR Lyrae and no ACs.
Even if  these could be underestimates,  the number of possible contaminants from M31 appears to be in any case  
small compared to the 31 RR Lyrae stars and 8 ACs we have 
found in And~XIX.
On the basis of the arguments above, we conclude that the contamination from M31 
  halo does not affect significantly our results.
%but we are aware that the sigma of the parameters derived can be larger given the lower statistic.

\section{SPATIAL DISTRIBUTION}\label{spatial}

%{\it The position of the variable stars in the LBC field is shown in Figure~\ref{fig:map}. 
%The black dots, which magnification depends on the $V$-mag, are all the objects plotted in the 
%CMD in the right panel of Figure~\ref{fig:dia}.
%The RR Lyrae stars are marked with  red circles and blue pentagon while  the  ACs are represented with green squares. 
%The yellow ellipse was drown using the half-light radius, ellipticity
%and position angle of  And~XIX given by \citet{mac08}.}

\begin{figure*}
\centering
\includegraphics[trim=0.001mm 5cm 0.001mm 8cm, keepaspectratio=true, scale=0.75]{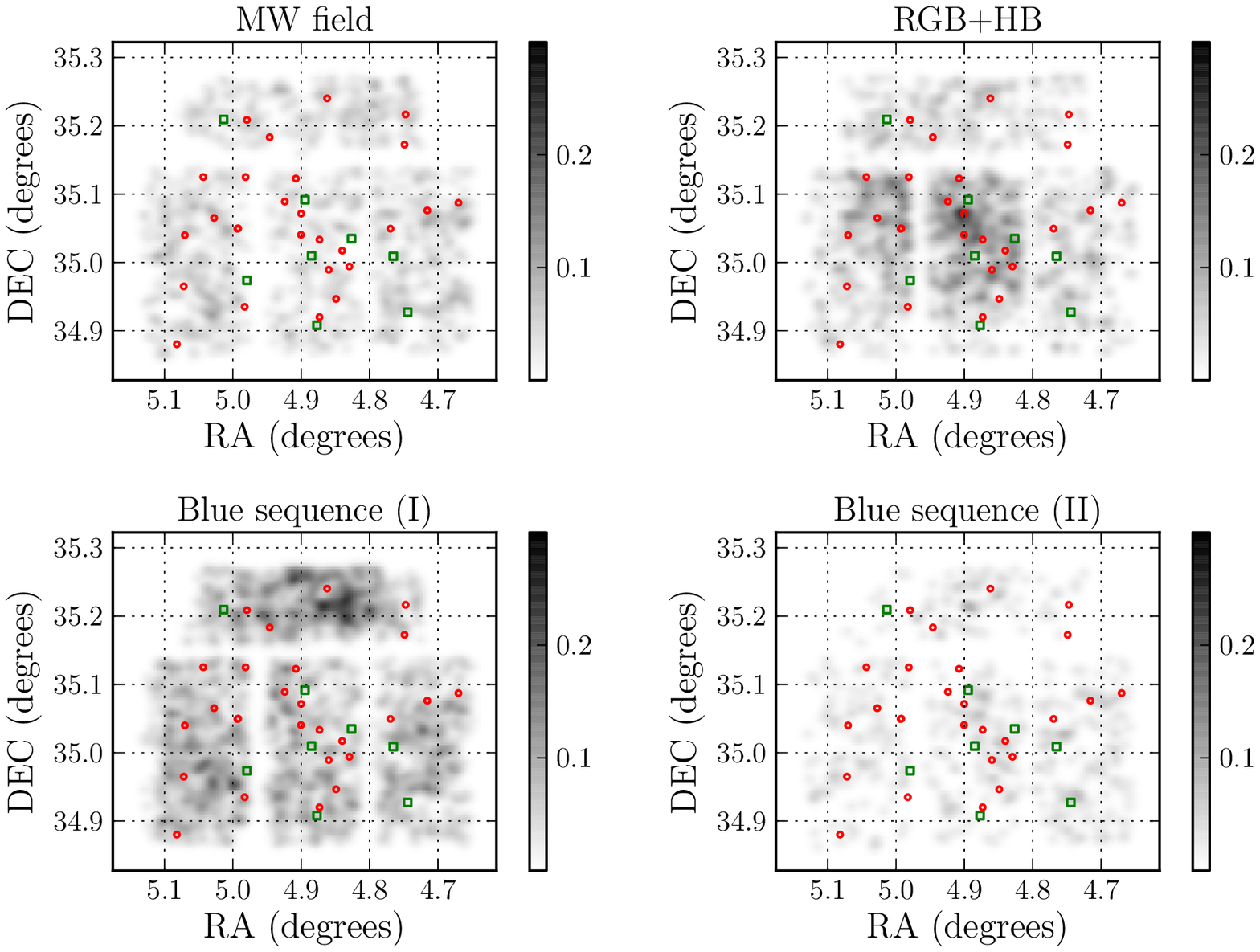}
\caption[]{Density-contours of the objects selected using the CMD in
  Figure~\ref{fig:cmdsel}. From the upper left panel clockwise MW stars (in
  green in Figure~\ref{fig:cmdsel}), RGB+HB stars (in magenta in
  Figure~\ref{fig:cmdsel}), intermediate blue sequence objects (in
  blue in Figure~\ref{fig:cmdsel}) and blue sequence objects (in cyan
  in Figure~\ref{fig:cmdsel}). Red circles are RR Lyrae stars
  while green squares are ACs. }
\label{fig:densmap}
\end{figure*}

\begin{figure*}
\centering
\includegraphics[keepaspectratio=true, scale=0.62]{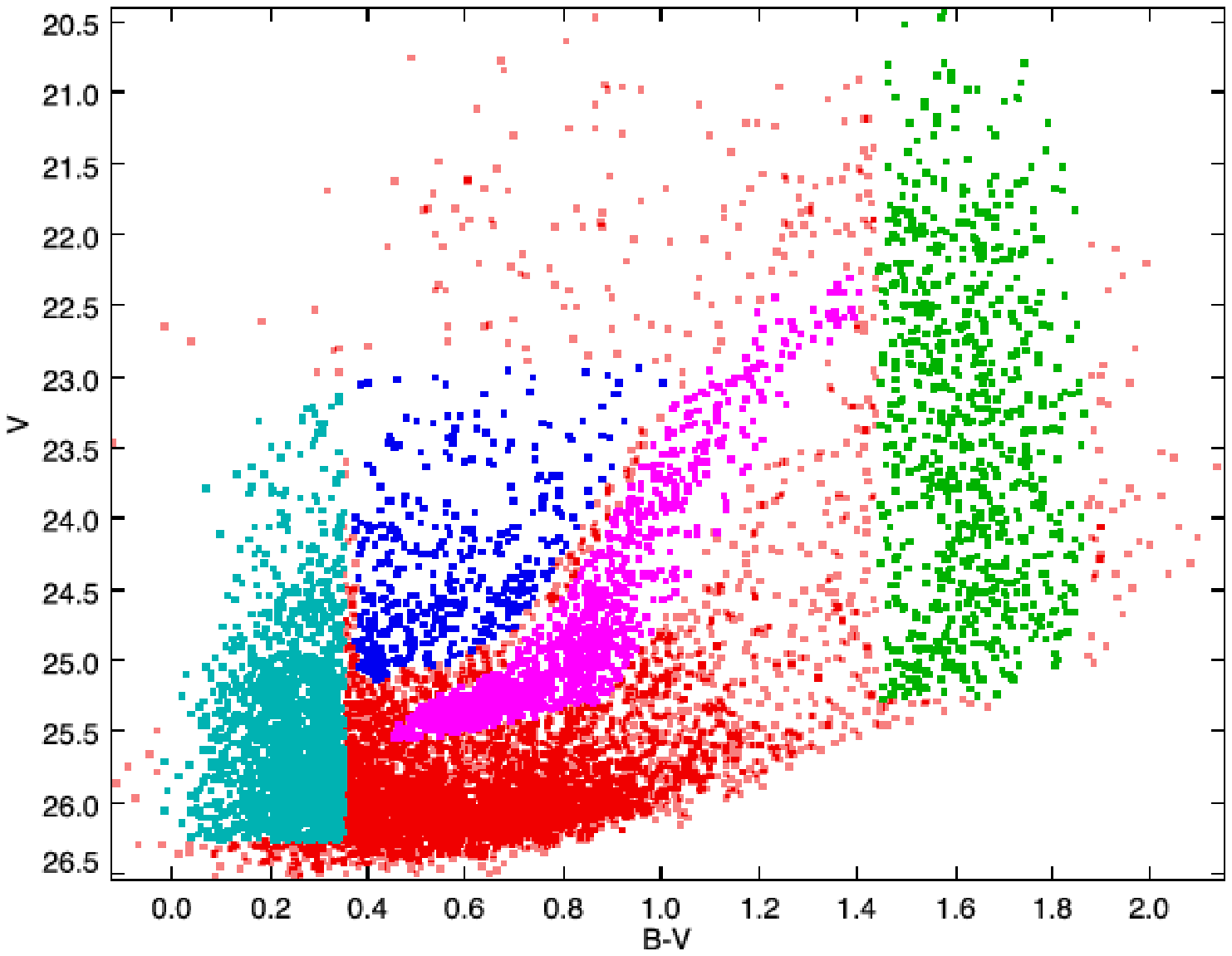}
\caption[]{CMD of the sources in the FoV of our observations of And~XIX satisfying the $\chi$ and Sharpness conditions described in Sect. 6 (red points),   with marked in different colours  the selections used to construct the spatial maps of
  Figure~\ref{fig:densmap}: RGB+HB stars (magenta), MW stars (green),
  blue objects (cyan) and intermediate blue objects (blue).}
\label{fig:cmdsel}
\end{figure*}

In order to explore the spatial structure of And~XIX, 
Figure~\ref{fig:densmap} shows the spatial distributions of stars from different regions 
 of the CMD (Fig.~\ref{fig:cmdsel}). To better visualize the maps, the data points were binned to a pixel size 
of 7 arcsec and smoothed with a 2D Gaussian kernel with $\sigma$=20 arcsec.
In Figure~\ref{fig:densmap} from the top left panel clockwise we show MW 
(green symbols in Fig.~\ref{fig:cmdsel}) stars, And~XIX RGB and HB stars (magenta 
symbols in Fig.~\ref{fig:cmdsel}), intermediate color objects (blue symbols in Fig.~\ref{fig:cmdsel}) 
and blue objects (cyan symbols in Fig.~\ref{fig:cmdsel}). Variable stars (red circles  for RR Lyrae stars, and green squares for
ACs) are also overlaid to the density maps.

\begin{figure}[]
\centering
\includegraphics[keepaspectratio=true, scale=0.4]{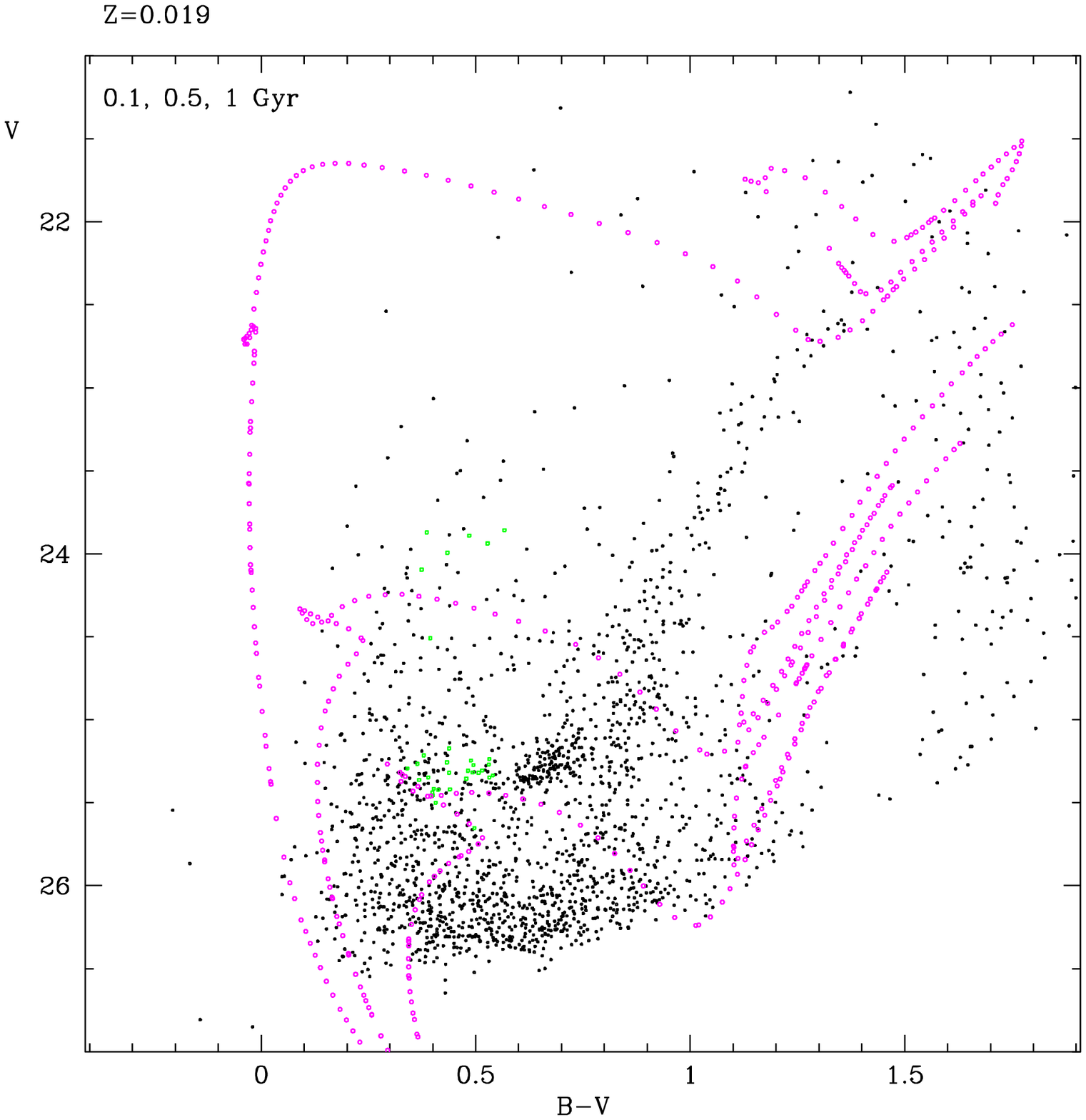}
\caption[]{CMD with overlaid  solar metallicity isochrones from 0.1 to 1 Gyr.}
\label{fig:solar}
\end{figure}

As expected, the MW stars are homogeneously distributed all over the field. On the
other hand, RGB and HB stars seem to be concentrated along a diagonal
bar-like structure running from southwest to northeast and  pointing
toward the M31 center. Along this bar are also positioned 20 of the
39 variable stars.
%In particular, both RR Lyrae and RGB/HB samples show similar concentration with 75$\%$ 
%of object within the half-light radius while ACs are apparently 
%more concetrated with only 10$\%$ 
%beyond the half light radius (although low number statistics do not allow a firm conclusion). 
Interestingly, the distribution of the blue objects does 
not correlate with the distribution of RGB and HB stars, which are clearly 
 members of And~XIX, but shows an over-density in the upper CCD 
 of the LBC camera, in a region around  R.A.=4.84$^{\circ}$, dec$=+35.22^{\circ}$, 
  and radius $\sim$2.5 arcmin. We suggest that these blue objects
are most likely unresolved galaxies. To examine in depth this possibility
we calculated an upper limit for the number of unresolved galaxies
expected in the FoV of our  LBC observations using the HST Ultra Deep Field (UDF)
catalog of galaxies by \citet{coe06} and making the assumption that
the distribution of galaxies in the sky is almost isotropic.  In the
UDF catalog we selected galaxies with the same range of colors and
magnitudes used to select the blue objects in our CMD. Furthermore, we
selected galaxies with radii smaller than 0.75 arcsec, which, given
the average seeing of the LBC images, should result in unresolved
objects. After scaling for the different area surveyed by the UDF
catalog of \citet{coe06} and by the LBC, we end up with an estimated
upper limit of 1450 unresolved galaxies in the LBC field.
The number of blue objects in the LBC catalog is 1736 that is
comparable to the upper limit found from the UDF catalog.  This seems to support our claim 
 that the majority of the blue objects in the CMD are likely
unresolved galaxies. As a further evidence Fig.~\ref{fig:solar} shows isochrones
with solar metallicities overlaid to the CMD. 
This figure shows that the blue  sources in the CMD (in cyan in Fig.~\ref{fig:cmdsel}) are much redder than
solar isochrones, so more consistent with unresolved galaxies than
single stars. In this case the overdensity in the upper CCD is likely
a cluster of galaxies. We searched the \citet{wen13} catalog  for known clusters of galaxies 
around the center of the overdensity ($\sim$ R.A.= 4.84$^{\circ}$, dec=35.22$^{\circ}$),
but we did not find any in a radius of  $\sim$2.5 arcmin. 
However,  we notice that \citet{wen13} catalog  is based on SDSS data, that are
 shallower when compared to our deep photometry.
 
Finally, the distribution of intermediate color objects partially
overlaps with the RGB/HB distribution (which is peaked in the central 
CCD), but shows mild overdensities in the upper and right hand CCDs
as well. This sample is likely mostly populated by unresolved
galaxies, although there are probably also some members of And~XIX. Indeed the ACs
are in this region of the CMD, and are contributing to these
overdensities. However, only high resolution observations with HST will
allow a better galaxies/stars separation, thus helping in clarifying whether 
And~XIX hosts an intermediate/young age component.

\section{CMD INTERPRETATION}

\subsection{Old population: two episodes of star formation}\label{sec:oldpop}

\begin{figure*}[!t]
\centering
\includegraphics[scale=0.9]{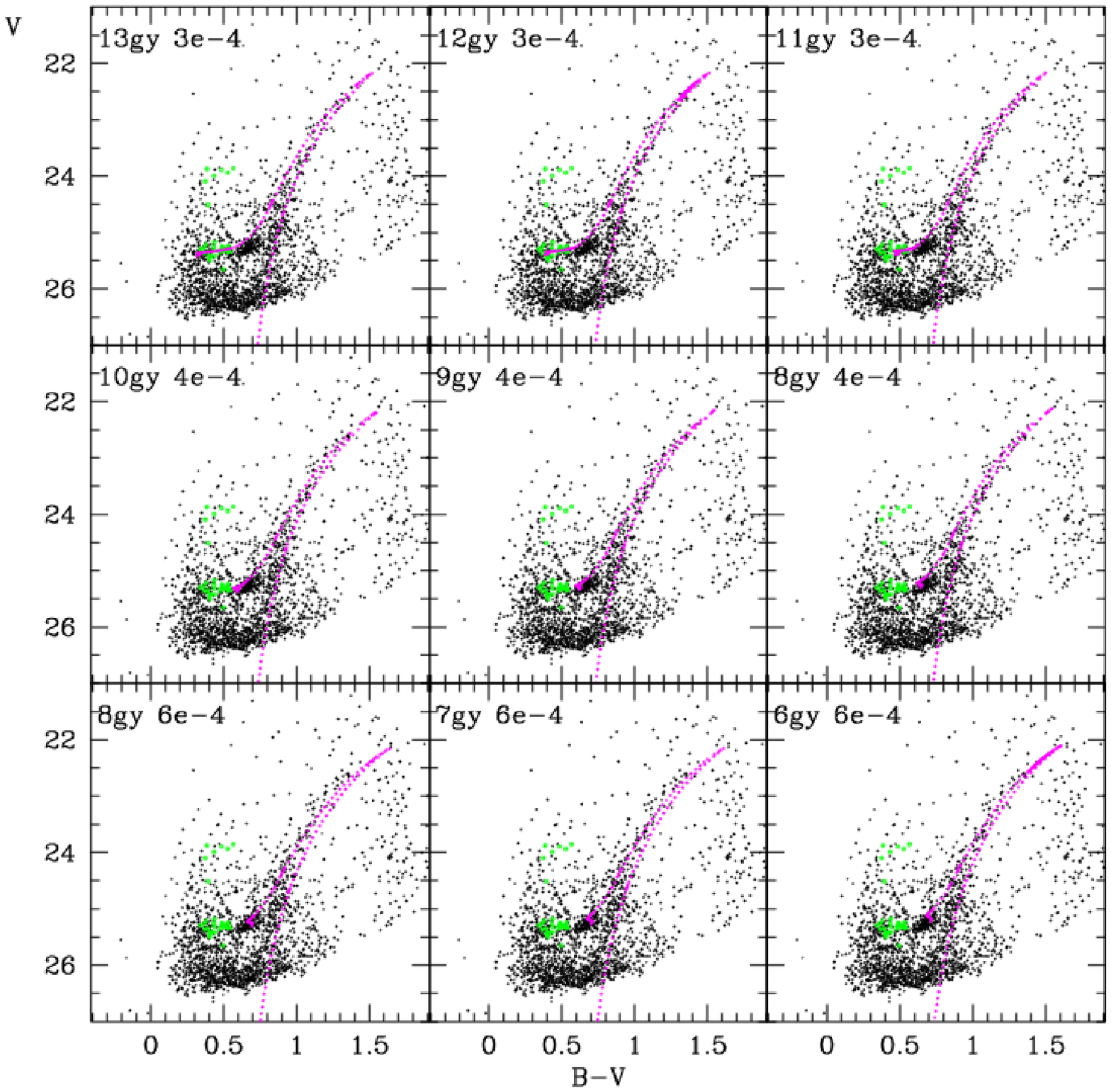}
\caption[]{Observed CMD with overlaid stellar isochrones from the Padova evolutionary models on the CMD 2.5 web interface. Metallicities from the top to the bottom panel 
are Z=0.0003, 0.0004 and 0.0006, respectively. The ages of the isochrones are indicated in each individual panel.
Variable stars are represented by green squares.}
\label{fig:old}
\end{figure*}

Overlaid to the CMD in Figure \ref{fig:old}  are shown the Padova isochrones of different ages and metallicities 
obtained using 
the CMD 2.5 web inferface\footnote{$http://stev.oapd.inaf.it/cgi-bin/cmd$} based on models from \citet{bres12}. The adopted foreground reddening and the distance
modulus are E($B-V$)=0.11 mag and $(m-M)_0=24.52$ mag, respectively, as derived from the RR Lyrae stars. Although the
age-metallicity degeneracy makes the interpretation of the RGB color
troublesome, we can still find feasible scenarios by fitting the RGB and HB
simultaneously. Our findings suggest that And~XIX hosts two distinct
stellar populations:

\begin{enumerate}
 \item  An old and metal poor component, hereinafter P1, as suggested by the
presence of RR Lyrae stars. Our fit suggests that only  metallicity
$Z=0.0003$ ($\log  Z/Z_{\odot}= -$1.80) and ages between 12 and 13 Gyr can fully match the location
of the RGB and the average color of the RR Lyrae stars. Formally, we can rule
out younger ages (the top right panel illustrates a 11 Gyr isochrone),
since the corresponding RGB and HB are too blue and too red
respectively, and higher metallicities ($Z=0.0004$, $\log Z/Z_{\odot}= -$1.67; middle panels), since
the predicted HB is clearly redder than the RR Lyrae color.

\item A more metal rich and possibly younger component, hereafter P2, as
traced by the red HB. Based on the isochrone fitting, as shown in middle
and bottom panels of Figure \ref{fig:old}, we find that isochrones in the
metallicity range $Z=0.0004-0.0006$ ($\log  Z/Z_{\odot}= -$1.67 / $-$1.5), with ages spanning from 6 to 10
Gyr, match well both the color extension of the red HB and the mean position
of the RGB. However, the lowest metallicity isochrones actually fit best,
while the more metal rich isochrones predict giants that are too red.
Lowering the age of the isochrone ($<\,6$ Gyr) partially counters this
effect, but also produces a too bright HB. Vice versa, isochrones older
than 10 Gyr at $Z=0.0004$, although producing tolerably good fits in the
RGB region, miss the red HB.

\end{enumerate}

Unfortunately, the strong galactic contamination makes it difficult to
quantify the fraction of blue HB stars. Likewise, the ratio between P1
and P2's star formation rates is very uncertain.

\subsection{A recent episode of star formation?}\label{sec:recent}

\begin{figure*}[!t]
\centering
\includegraphics[scale=0.9]{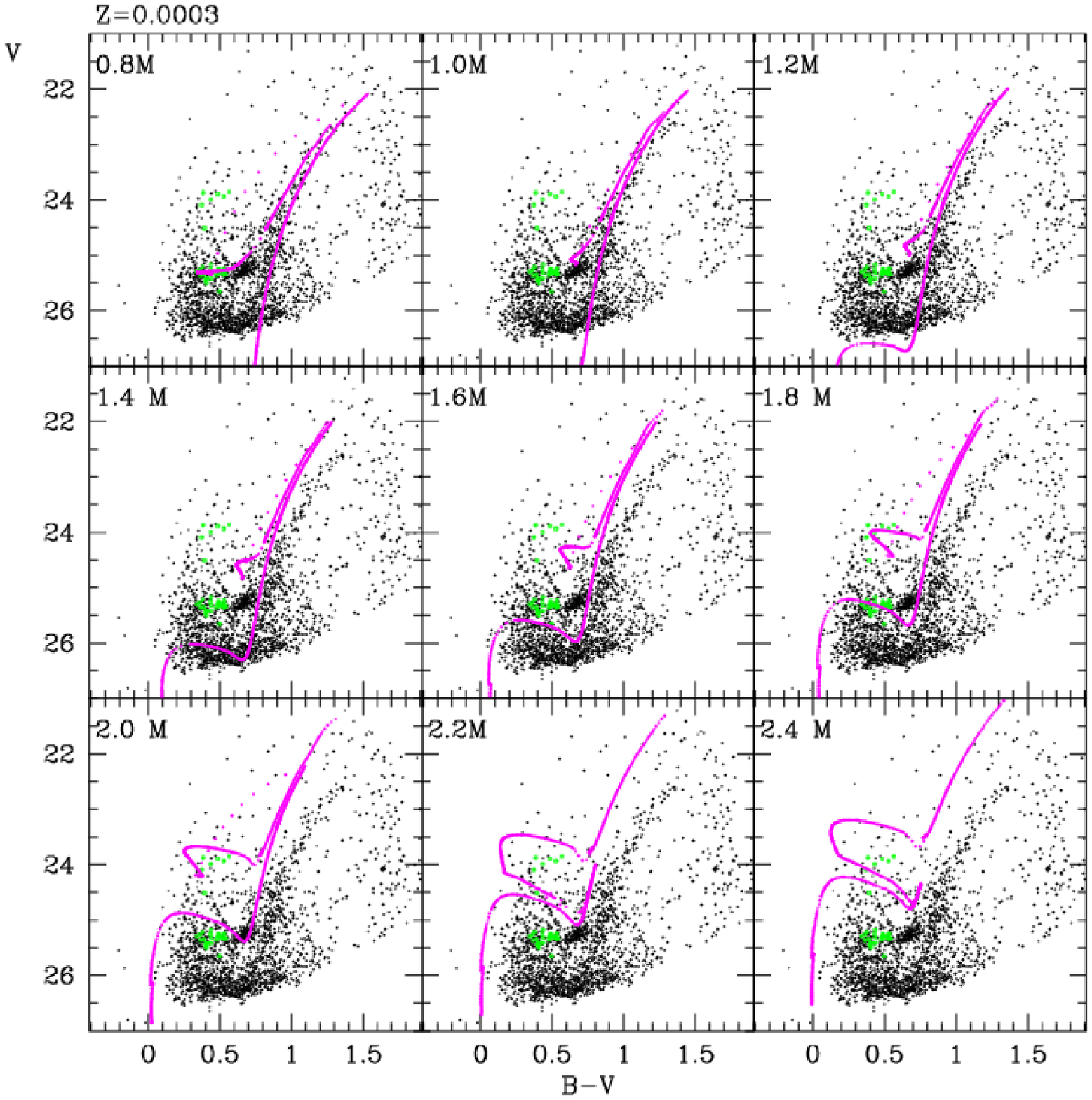}
\caption[]{Stellar evolutionary tracks from the Basti web site. Starting from the top-left to the bottom-right masses from 0.8 to 
2.4 M$_\odot$. The metallicity is $Z=0.0003$. Variable stars are represented with green squares.}
\label{fig:trac03}
\end{figure*}

\begin{figure*}[!t]
\centering
\includegraphics[scale=0.9]{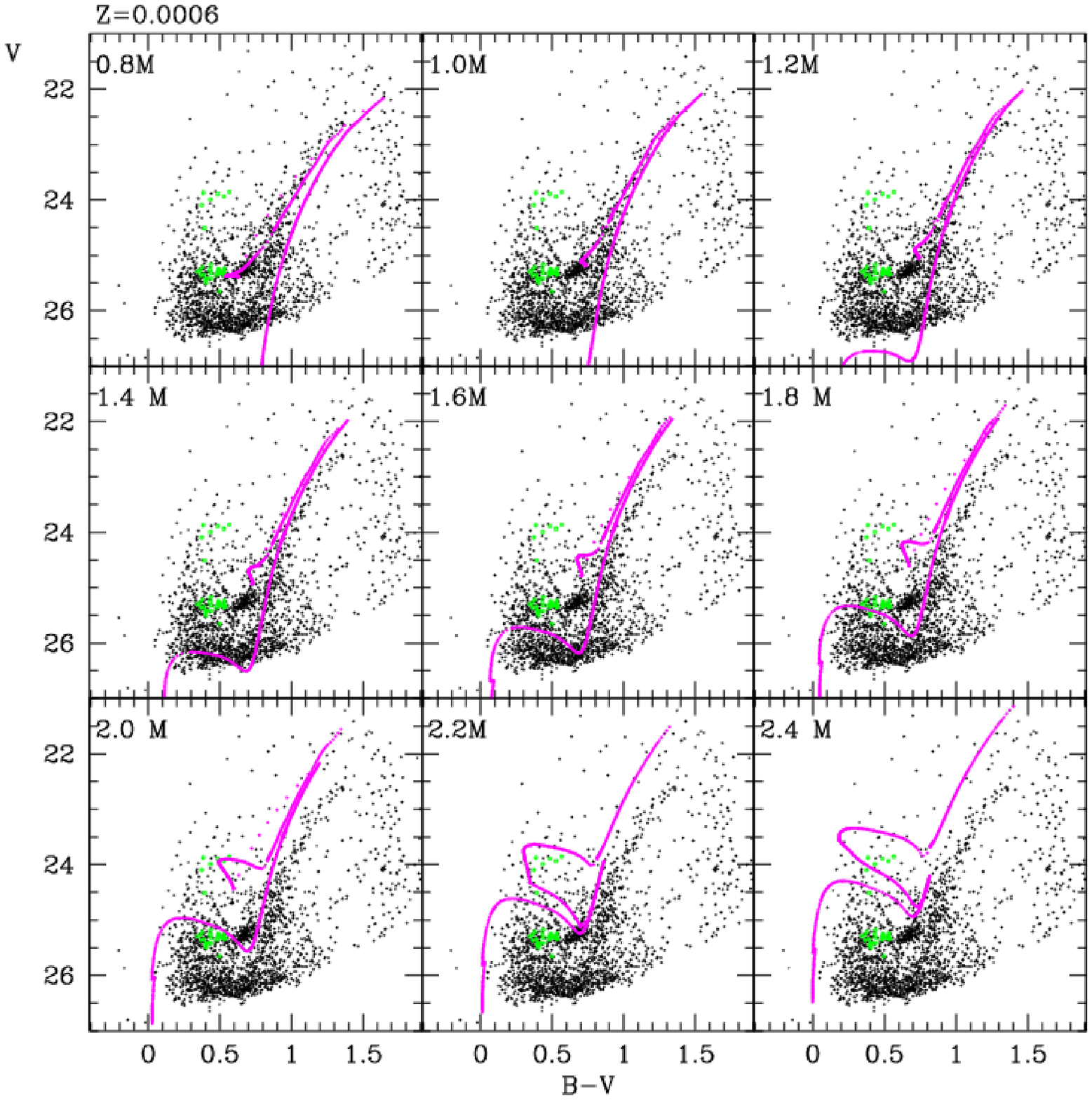}
\caption[]{Same as Figure.~\ref{fig:trac03}, but for metallicity $Z=0.0006$.}
\label{fig:trac06}
\end{figure*}

\begin{figure*}[!t]
\centering
\includegraphics[scale=0.9]{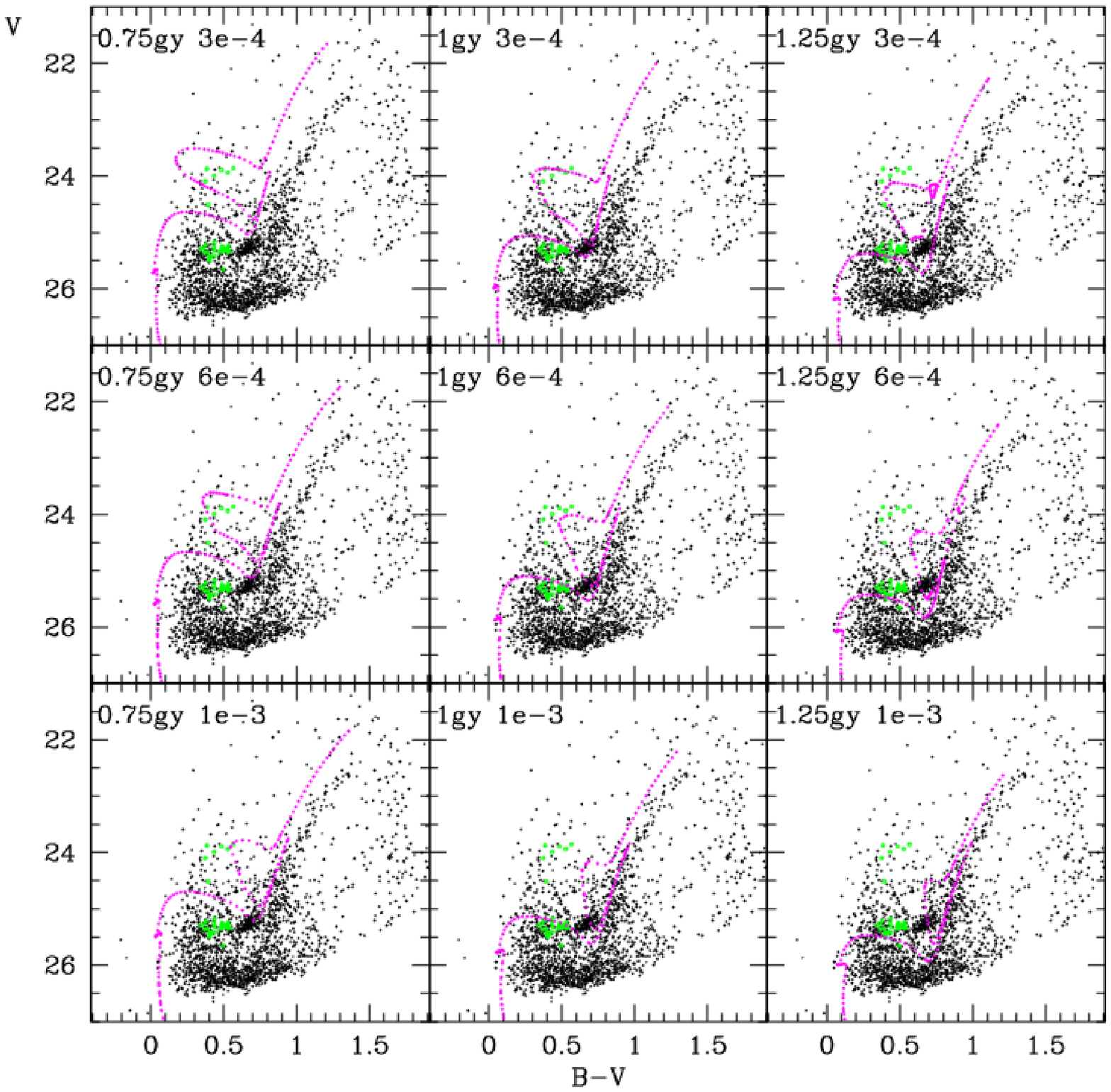}
\caption[]{Same as Figure \ref{fig:old}, but for metallicities from the top to the bottom panel 
of Z=0.0003, 0.0006 and 0.001, respectively and  younger isochrones.}
\label{fig:young}
\end{figure*}

The  presence of pulsating stars %about 1 mag
 brighter than RR Lyrae stars
rises the question of whether  And~XIX has been forming stars up to 1 Gyr ago. In
Figure \ref{fig:trac03} and \ref{fig:trac06} we show stellar evolutionary tracks from
the Basti web-site\footnote{$http://albione.oa-teramo.inaf.it/$} based on models by \citet{piet04}
for metallicities $Z=0.0003$ and
$Z=0.0006$ respectively, and masses in the range 0.8-2.4 M$_\odot$, overlaid on the observed 
CMD.  We used the  Basti  tracks for this comparison because the $Z=0.0003$ and
$Z=0.0006$ metallicities are not available for the Padova evolutionary tracks.
In the $Z=0.0003$ case, the location of the 
pulsating stars is bracketed by 1.8 and 2.0 $M_{\odot}$ tracks, while in
the $Z=0.0006$ case it is constrained by 2.0 and 2.2 $M_{\odot}$ tracks.
In terms of age, the best fitting isochrones (see Figure~\ref{fig:young}) suggest a
range of ages 1-1.25 Gyr old at $Z=0.0003$ and 0.75-1 Gyr old at
$Z=0.0006$. Although this is not a large difference, we note that in the
former scenario the pulsating stars are consistent with being ACs,
 as suggested by the RGB tip of similar luminosity for both the
1.8 and 2.0 $M_{\odot}$ tracks (both masses are lower than the RGB
transition mass), while in the latter they are at the borderline between
being ACs and Short-Period Classical Cepheids, as
suggested by the short RGB of the 2.2 $M_{\odot}$ track compared to the
2.0 $M_{\odot}$ track (the 2.2 $M_{\odot}$ is above the RGB transition
mass, the 2.0 $M_{\odot}$ is below). Furthemore, the stellar evolutionary tracks 
predict the existence of ACs only for metallicities  lower than $Z=0.0004$ (see, e.g.,  Marconi et al. 2004).

\begin{figure*}[]
\centering
\includegraphics[width=\textwidth,height=\textheight,keepaspectratio]{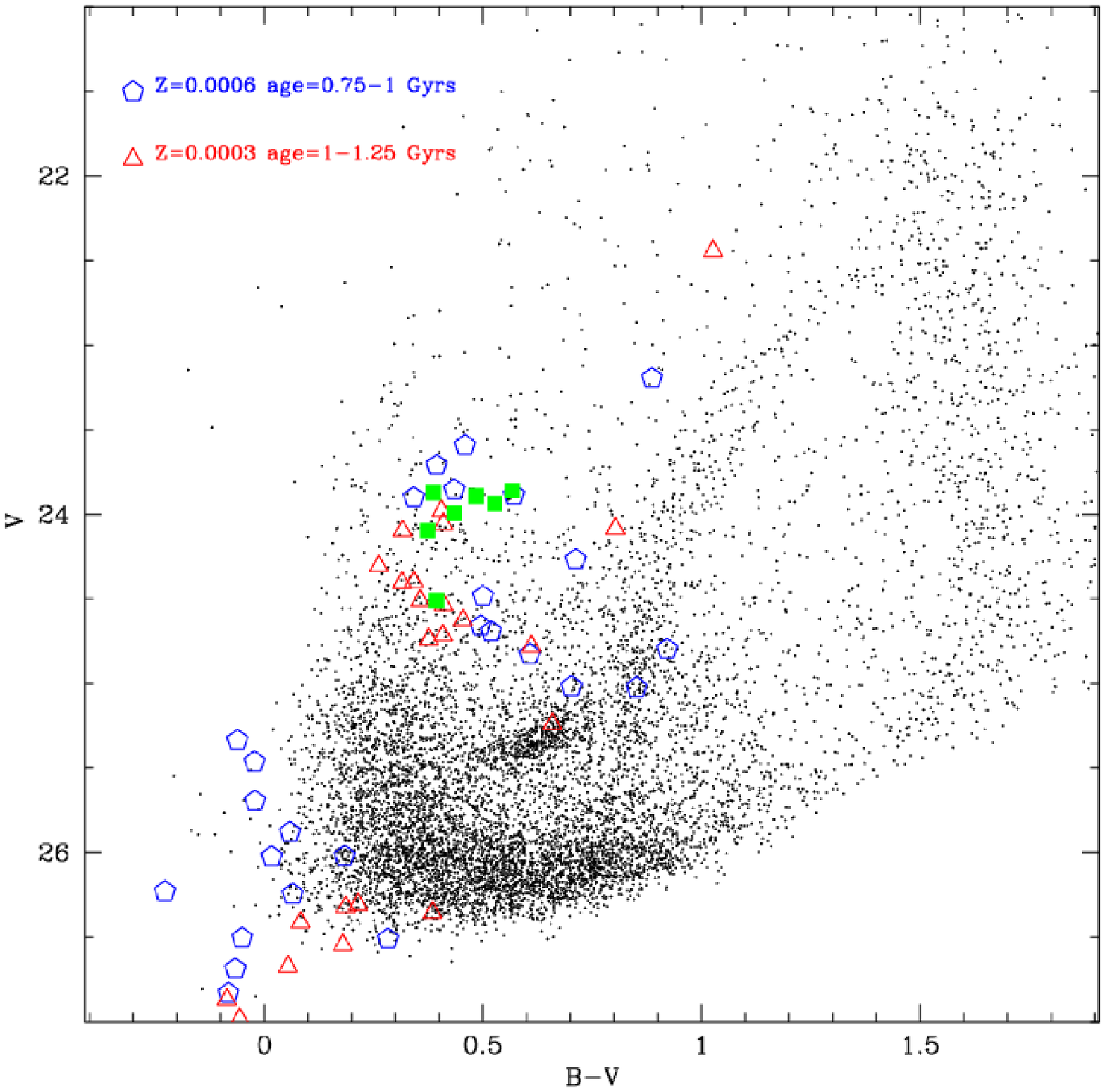}
\caption[]{Synthetic stellar population allowing to reproduce the number of ACs (green squares) on the observed CMD (see Section~\ref{sec:recent}
for details).}
\label{fig:sint}
\end{figure*}

To further investigate the presence of young stars, two synthetic 
populations \citep[see ][ for an overview of the technique]{cign10} 
were generated following the two most likely scenarios, namely a 
metallicity Z=0.0006 and a constant SF in the range 1-0.75 Gyr,  and a 
metallicity Z=0.0003 and a constant SF in the range 1.25-1.00 Gyr.  
For  both models we used the Padova tracks \citep{mari08,bert09}, 
convolved with photometric errors and incompleteness as 
estimated from artificial star tests, and a Salpeter initial mass function 
(IMF). Once the number of synthetic objects populating the region 
$23.5<V<24.5$ mag and $0.2<B-V<0.9$ mag equals the number of brighter pulsators, the procedure 
was stopped, giving the minimum amount of star formation necessary to 
generate the brighter variables. This led to a star formation rate of the order of 
$10^{-5}\,M_{\odot}$ per yr. Figure \ref{fig:sint} shows the resulting simulations 
(blue pentagons and red triangles indicate Z=0.0006 and Z=0.0003 
simulations, respectively) overlaid to the observed CMD. Although the exact 
CMD morphology of the ACs is not perfectly reproduced, both synthetic 
populations show a number of MS stars (objects at $V>25.5$ mag) which is not 
much lower than star counts observed in the corresponding CMD regions. 
This suggests that, if ACs are associated with a genuine SF episode, our 
inferred rate can be considered an upper limit to the recent activity in 
And~XIX.

The other way of accommodating the apparent youth is that these stars
are the evolved counterpart of MS blue stragglers. In this case they are
not the result of a recent episode of star formation, but rather the
result of mass transfer in close binary systems occurred about 1 Gyr
ago. Unfortunately, the detection of MS blue stragglers or genuine
young stars in MS is greatly hindered by the contamination of blue
galaxies.

\section{SUMMARY AND CONCLUSIONS}

We have presented $B$ and $V$ time-series observations of the M31 dSph satellite And~XIX, that we 
 performed using the  LBC at the  LBT.  A total number of 39 variable stars were identified in the galaxy of which 31 are RR Lyrae stars and 8 likely are  ACs.
From the average period of the RRab stars and the period-amplitude  diagram we  classify And~XIX as an Oo-Int system.
The average $V$ magnitude of the RR Lyrae stars allowed us to  estimate the distance modulus of And~XIX,  (m-M)$_0$=$24.52\pm0.23$ mag (for 
E(B-V)=0.11$\pm$0.06 mag, as we derive from the RR Lyrae stars) or (m-M)$_0$=$24.66\pm0.17$ mag (for E(B-V)=0.066$\pm$0.026 mag as derived from \citealt{sch98} maps).
Both estimates are in good agreement with the value of (m-M)$_0=24.57^{+0.08}_{-0.43}$ mag found by \cite{con12}.
Comparing the observed CMD with stellar isochrones we find evidence for two different stellar populations in And~XIX. 
One mostly made by old (13 Gyrs) and metal poor ([Fe/H]=-1.8 dex) stars which produced the RR Lyrae variables, and the second composed
by metal enriched stars (-1.5$<$[Fe/H]$<$-1.7 dex) with ages between 6 and 10 Gyrs.  
The presence of ACs in And~XIX provides hints for a recent episode of star formation in this galaxy.  
With the use of evolutionary tracks and isochrones, we constrained this formation episode 
in a epoch between 0.75 and  1.25 Gyrs ago.
% The spatial distribution of the RGB stars gives indication of a possible bar elongated toward the direction
%of the M31 center.
%The ACs found well follow a $PW$ relation. 
%The specific frequency of ACs in And~XIX is consistent with the one found in dSph galaxies.
The ACs are found to follow well a $PW$ relation and are mostly a genuine 
  population belonging to And~XIX with no or very little contamination by the 
  M31 halo. The specific frequency of ACs in And~XIX is also consistent with the
  value typical of other dSph galaxies in M31 and the MW.
 Finally, the spatial distribution of the RGB and HB 
  stars gives indication of the presence of a bar-like structure elongated
  in  the direction of the M31 center.

\acknowledgments

We warmly thank P. Montegriffo for the development and maintenance of the GRATIS software,  G. Beccari for useful 
suggestions on the estimate of the photometric completeness with ALLFRAME, Flavio Fusi Pecci, Monica Tosi  and Carla Cacciari 
for useful  comments and discussions on an earlier version 
of the paper.
Financial support for this research was provided by COFIS ASI-INAF I/016/07/0 and by PRIN INAF 2010 (P.I.: G. Clementini).
The LBT is an international collaboration among institutions in the United States,
Italy and Germany. LBT Corporation partners are: The University of Arizona o
n behalf of the Arizona university system; Istituto Nazionale di Astrofisica, Italy;
LBT Beteiligungsgesellschaft, Germany, representing the Max-Planck Society,
the Astrophysical Institute Potsdam, and Heidelberg University; The Ohio State
University, and The Research Corporation, on behalf of The University of Notre Dame,
University of Minnesota, and University of Virginia.
We acknowledge the support from the LBT-Italian Coordination Facility for the
execution of observations, data distribution and reduction.

%% To help institutions obtain information on the effectiveness of their
%% telescopes, the AAS Journals has created a group of keywords for telescope
%% facilities. A common set of keywords will make these types of searches
%% significantly easier and more accurate. In addition, they will also be
%% useful in linking papers together which utilize the same telescopes
%% within the framework of the National Virtual Observatory.
%% See the AASTeX Web site at http://www.journals.uchicago.edu/AAS/AASTeX
%% for information on obtaining the facility keywords.

%% After the acknowledgments section, use the following syntax and the
%% \facility{} macro to list the keywords of facilities used in the research
%% for the paper.  Each keyword will be checked against the master list during
%% copy editing.  Individual instruments or configurations can be provided 
%% in parentheses, after the keyword, but they will not be verified.

{\it Facilities:}  \facility{LBT}.

\end{document}